# Enhanced piezoelectric response of AlN via alloying of transitional metals, and influence of type and distribution of transition metals

*Xian-Hu Zha, Xiufang Ma, Jing-Ting Luo, and Chen Fu*[*]

X.-H. Zha, X. Ma, J.-T. Luo, C. Fu
Key Laboratory of Optoelectronic Devices and Systems of Ministry of Education and Guangdong Province, College of Physics and Optoelectronic Engineering, Shenzhen University, Shenzhen 518060, China.
Email: chenfu@szu.edu.cn.

**Keywords:** aluminum nitride, transition metal alloy, piezoelectric constant, type and distribution, density functional calculation

Aluminum nitride (AlN) is an important piezoelectric material for a wide range of applications, many efforts are devoted to improving its piezoelectric response by alloying with transition metals (TMs). In this paper, the influence of the type and distribution of TM on the piezoelectric response is discussed for the first time. $TM_{0.0625}Al_{0.9375}N$ with twenty-eight different TMs are investigated, and most show higher values of piezoelectric strain modulus $d_{33}$ than that of AlN. This is because the TM introduces weaker TM-N bonds and locates closer to the centre of three neighbouring N atoms. The location of TM is determined to be significantly correlated with its group number. Alloys of $TM_xAl_{1-x}N$ (TM=Sc, Cr, Sr, Mo, Ru and Rh) with varying $x$ are further studied. On basis of the cost of the TMs and piezoelectric performances, the alloy with Mo is more effective in enhancing $d_{33}$. A high $d_{33}$ of 12.3 times that of pure AlN is realized in a metastable configuration of $Mo_{0.167}Al_{0.833}N$. The distribution of Mo plays a key role in the piezoelectric performance. A higher $d_{33}$ is more likely to appear in $Mo_xAl_{1-x}N$ with more Al sublayers containing Mo atoms and with fewer dimers of Mo atoms along the $z$-axis.

## 1. Introduction

Piezoelectric materials are essential layers in microelectromechanical-based systems (MEMSs),[1, 2] with a wide range of applications in surface and bulk acoustic resonators,[3] radio frequency (RF) signal processing,[4] biosensing,[5] among others.[1, 2] A large number of piezoelectric materials have been reported, such as quartz ($SiO_2$),[6, 7] lead zirconate titanate (PZT), lithium niobate ($LiNbO_3$), zinc oxide (ZnO), and aluminum nitride (AlN).[2, 8] Due to its advantages of stability at high temperature, excellent thermal conductivity, strong mechanical strength, non-toxic, and CMOS compatibility,[9] AlN has been widely used in MEMSs, especially for devices operating in harsh environments, such as high power[3] and high temperature.[10, 11] However, the piezoelectric response of AlN is lower than that of many other common piezoelectric materials, with values of $d_{33}$ in the range of 4.50-6.40 *pC/N*. The corresponding value in ZnO is twice as high, and the value in PZT is dozens of times higher.[2, 12] Improving the piezoelectric constant of AlN has therefore become an important concern.

Alloying with transition metals (TMs) has been demonstrated to be effective in improving the piezoelectric constant of AlN. Akiyama *et al.* first synthesized a $Sc_xAl_{1-x}N$ alloy by dual reactive co-sputtering, and achieved a high $d_{33}$ value of 27.6 *pC/N* with $x$ equal to 0.430.[13-15] The enhancement in $d_{33}$ was explained by the increased response of internal coordinates to strain and the softening of the elastic constant $c_{33}$.[16] Compared to AlN, the Sc-alloy configuration was also determined to have a larger electromechanical coupling coefficient[17] and lower dielectric loss,[18] which enables its potential application in energy harvesting and sensors,[18] and wideband resonator filters.[19-21] However, Sc is an expensive and scarce TM and thus the Sc-alloy AlN is not suitable for large-scale applications. Moreover, it is difficult to obtain a uniform distribution of the Sc-alloy configuration at high alloy content because of its preferential phase separation into rocksalt ScN and wurtzite AlN phases.[2, 12] Furthermore, other TMs as well as B have been adopted for incorporation in AlN. The introduction of B[22, 23] enhanced the hardness of AlN, and the piezoelectric constant showed significant variation with the B coordinates. A $Y_xAl_{1-x}N$ thin film was realized by reactive magnetron sputtering,[24] which exhibited a higher $d_{33}$ value compared to AlN.[25] Co-alloying of AlN with BN and YN has also been investigated theoretically, and both the elastic and piezoelectric properties were found to improve.[26] A $d_{33}$ value of 9.41 *pC/N* was obtained in an $Er_xAl_{1-x}N$ film, which is almost twice that of AlN.[27] Ta-doping was found to facilitate the *z*-axis orientation of AlN, and the $d_{33}$ of the Ta-doped AlN increased to 8.20 *pC/N* at an alloy content of 0.0510.[28] Alloying with Yb in AlN enhanced the electromechanical coupling coefficient and piezoelectric constant,[29, 30] and a maximum $d_{33}$ value of 12.0 *pC/N* was determined for $Yb_{0.300}Al_{0.700}N$. $Cr_xAl_{1-x}N$ is also attracting increasing attention because its $d_{33}$ is even higher than that of the Sc-alloy configuration at a low value of $x$.[12, 31, 32] Moreover, alloys of multiple TMs in AlN have been investigated theoretically, such as Mg with X (X=Nb, Ti, Zr, Hf, Cr, Mo and W)[33-37] and Li with X (X=V, Nb and Ta).[38]

Although several TMs have been incorporated in AlN, various other TMs are yet to be investigated, and it is still desirable to find an economical and effective approach to improve the piezoelectric constant of AlN. The enhancement of $d_{33}$ in TM-alloy AlN has been explained in terms of the increased response of internal coordinates to strain and the decreased elastic constant,[16, 36] but it might be more productive for experimental scientists to establish the relationships between the piezoelectric constant and the type and distribution of the alloying element. Regarding these issues, the piezoelectric performances of AlN alloyed with a series of TMs (TM=Na, Mg, K, Ca, Sc, Ti, V, Cr, Mn, Fe, Co, Ni, Cu, Zn, Ga, Rb, Sr, Y, Zr, Nb, Mo, Tc, Ru, Rh, Pd, Ag, Cd and In) denoted as $TM_{0.0675}Al_{0.9325}N$ are investigated in this work. Most of the TM-alloy configurations show higher $d_{33}$ than that of pure AlN. The values of $d_{33}$ for $TM_{0.0675}Al_{0.9325}N$ (TM=Sr, Mo, Ru and Rh) are even higher than those of the well-studied Sc- and Cr-alloy configurations at the same alloy content. The thermodynamic stability and piezoelectric properties of $TM_xAl_{1-x}N$ (TM=Sc, Cr, Sr, Mo, Ru and Rh) with values of $x$ up to 0.250 are further investigated. On basis of the cost of TMs and piezoelectric performance, the Mo alloy is more effective in improving the $d_{33}$ of AlN. The influence of Hubbard U correction and supercell size on the thermodynamic stability and $d_{33}$ of $Mo_xAl_{1-x}N$ is also discussed. Based on different special quasirandom structures (SQSs), an average $d_{33}$ of 3.61 times that of AlN is realized in $Mo_{0.250}Al_{0.750}N$. A high $d_{33}$ of 58.6 $pC/N$ is determined in a metastable SQS of $Mo_{0.167}Al_{0.833}N$.

## 2. Results and discussion

Before investigating the TM-alloy configurations, pure AlN is studied first. The lattice parameters $a$ and $c$ of the hexagonal AlN unit cell are determined to be 3.13 and 5.01 $Å$ respectively. Corresponding $c/a$ and the internal parameter $u$ are found to be 1.60 and 0.382 respectively. Values of all of these lattice parameters are consistent with those found in previous studies,[12, 13, 18, 33] which implies that the computational parameters are reliable. Furthermore, the $2 \times 2 \times 2$ AlN supercell with an Al atom replaced by TM denoted as $TM_{0.0625}Al_{0.9375}N$ is studied. Top and side views of the 2×2×2 AlN supercell are provided in **Figures 1**a and 1b. The TMs studied are denoted in light blue in the periodic table of elements in Figure 1c. Values of $a$ and $c$ for these TM-alloy configurations are provided in Figure 1d. The variations in lattice parameters are generally in line with the changing trend of the atomic radii of the TM, as shown in **Figure S1** in the Supporting Information (SI). For TMs in the same group in the periodic table of elements, $a$ increases with the period number of the TM. With increasing group number, $c$ varies more significantly when the TM is in a larger period. For TMs in period IV, $a$ generally decreases to $Ni_{0.0625}Al_{0.9375}N$, and $c$ decreases to $Cr_{0.0625}Al_{0.9375}N$. The smallest $a$ and $c$ are 6.26 and 10.0 $Å$ respectively. Regarding TMs in period V, $c$ generally decreases with TMs from Rb to Ru, and then increases from Ru to Cd. The value of $c$ for $Mo_{0.0625}Al_{0.9375}N$ of 10.1 $Å$ is higher than those of neighbouring Nb- and Tc-alloy configurations. The slight difference between the change trends in $a$ and $c$ could be ascribed to the details of bonding behavior.[39] Based on the lattice parameters, values of $c/a$ for these TM-alloy configurations are calculated and are provided in **Figure S2**. The largest and smallest ratios are 1.62 and 1.59, for $Rb_{0.0625}Al_{0.9375}N$ and $Ru_{0.0625}Al_{0.9375}N$ respectively.

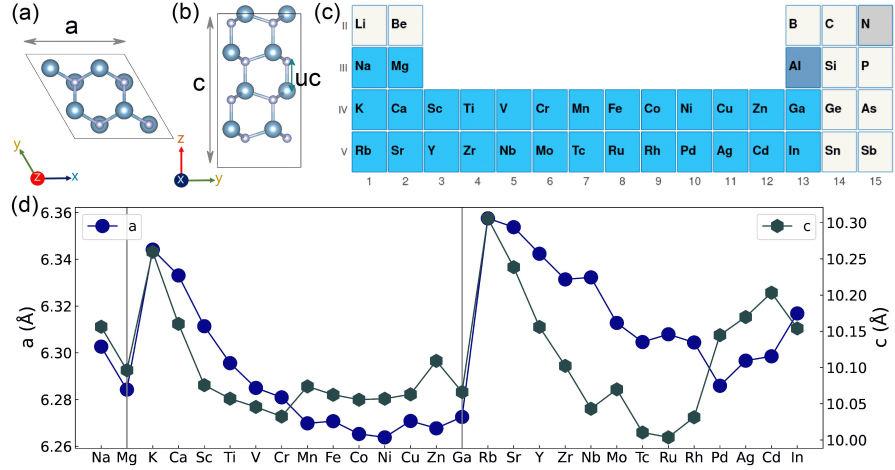

**Figure 1.** a) top- and b) side-views of the $2 \times 2 \times 2$ supercells of AlN. The dark blue, and small grey balls denote Al and N respectively; c) periodic table of elements, where the TMs studied are denoted in light blue; d) values of $a$ and $c$ of $TM_{0.0625}Al_{0.9375}N$, which are denoted in blue circles and dark green hexagons, respectively.

On the basis of the relaxed structures, piezoelectric stress ($e_{ij}$) and strain ($d_{ij}$) moduli are studied. Since the polarization of AlN is mainly along the $z$-axis, $e_{33}$ and $d_{33}$ are elaborated in detail. The value of $e_{33}$ is determined in two parts:[40]

$$e_{33} = e_{33\_clamped} + e_{33\_nonclamped} \qquad (1)$$

where $e_{33\_clamped}$ is the clamped-ion term implying the effect of strain on the electronic structure; and the second term, $e_{33\_nonclamped}$, shows the effect of internal strain on the polarization,[12, 40] which is evaluated as:

$$e_{33\_nonclamped} = \frac{4eZ_{33}^*}{\sqrt{3}a_0^2} \frac{du}{d\delta} \qquad (2)$$

where $a_0$ is the lattice parameter in the basal plane, $\dfrac{du}{d\delta}$ represents the response of the internal parameter $u$ to the strain $\delta = \Delta c / c_0$ applied along the $z$-axis, and $Z_{33}^*$ denotes the Born charge in the unit of $e$. With the piezoelectric stress modulus, the piezoelectric strain modulus is further calculated according to the following equation:

$$d_{ij} = \sum_{k=1}^{6} e_{ik} s_{kj} \qquad (3)$$

where $S_{jk}$ is the element of the elastic compliance matrix $S$. All of the subscripts in the piezoelectric and elastic constants are given in Voigt notation. According to the above equation, $d_{33}$ is calculated as:

$$d_{33} = e_{31}s_{13} + e_{32}s_{23} + e_{33}s_{33} + e_{34}s_{43} + e_{35}s_{53} + e_{36}s_{63} \qquad (4)$$

From **Equations 1** and **2**, $e_{33}$ and its $e_{33\_clamped}$ and $e_{33\_nonclamped}$ parts of $TM_{0.0625}Al_{0.9375}N$ are calculated and provided in **Figure 2**a, which shows that all values of $e_{33\_clamped}$ are negative. The smallest and largest values are -1.09 and -0.286 $C/m^2$ for $Zn_{0.0625}Al_{0.9375}N$ and $Cr_{0.0625}Al_{0.9375}N$ respectively. The value of $e_{33}$ is mainly contributed by the non-clamped ionic term, $e_{33\_nonclamped}$, and in $TM_{0.0625}Al_{0.9375}N$ (TM=Sc, Ti, V, Cr, Sr, Mo, Ru and Rh) the values are all higher than 2.00 $C/m^2$. Based on the contributions of the two parts, the $e_{33}$ of pure AlN is determined to be 1.40 $C/m^2$, which is denoted as a dashed line in Figure 2a for comparison. $Zn_{0.0625}Al_{0.9375}N$ shows the lowest value of -0.108 $C/m^2$. Values of $e_{33}$ for $TM_{0.0625}Al_{0.9375}N$ (TM=Ca, Sc, Ti, V, Cr, Fe, Co, Ga, Sr, Y, Nb, Mo, Ru and Rh) are higher than that of pure AlN, and the values for $TM_{0.0625}Al_{0.9375}N$ (TM=Cr, Mo, Ru and Rh) are even higher than 1.70 $C/m^2$.

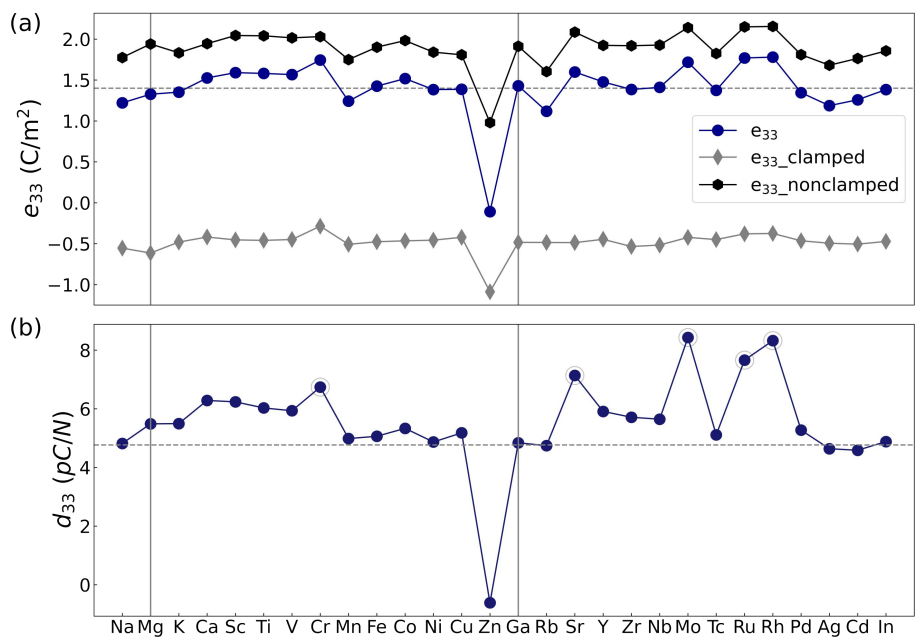

**Figure 2.** a) $e_{33}$ of $TM_{0.0625}Al_{0.9375}N$. The dark blue circles, grey diamonds, and black hexagons show $e_{33}$ and its $e_{33\_clamped}$ and $e_{33\_nonclamped}$ parts respectively; b) $d_{33}$ of $TM_{0.0625}Al_{0.9375}N$, where the circled points are values higher than or equal to the $d_{33}$ of $Cr_{0.0625}Al_{0.9375}N$. The grey dashed lines are corresponding values in pure AlN.

Values of $d_{33}$ calculated from **Equation 4** are shown in Figure 2b. The value for pure AlN is found to be 4.77 $pC/N$, which is close to previously reported values.[12, 25, 33] Unexpectedly, most TM-alloy configurations show higher $d_{33}$ values than that of pure AlN. The values of $TM_{0.0625}Al_{0.9735}N$ (TM=Sc and Cr) are calculated to be 6.24 and 6.74 $pC/N$ respectively, which are close to experimental values.[12, 13, 31] It is worth noting that values of $d_{33}$ for $TM_{0.0625}Al_{0.9375}N$ (TM=Sr, Mo, Ru and Rh) are higher than that of $Cr_{0.0625}Al_{0.9375}N$. The highest value of 8.43 $pC/N$ is for $Mo_{0.0625}Al_{0.9375}N$, which is 76.9% higher than that of pure AlN. All the other non-zero $d_{ij}$ values as well as $d_{33}$ are provided in **Table S1**. From the table, other $d_{ij}$ values of AlN also vary significantly after TM-alloying. Additionally, the value of $d_{33}$ has been estimated as $d_{33\_from\_c_{33}}=e_{33}/c_{33}$ in previous work,[18, 30] and values of $d_{33}$ from the two different approaches are compared in **Figure S3**. From the figure, the change trends of $d_{33}$ and $d_{33\_from\_c_{33}}$ are similar, although the latter slightly underestimates the value. The similar change trends imply that it is reasonable to use the variation of $c_{33}$ to estimate the influence of mechanical strength on $d_{33}$.

As in previous studies,[12, 16, 30, 34, 41] the increase in $d_{33}$ in TM-alloy AlN could be ascribed to the increase in $e_{33}$ and the softening of $c_{33}$. The increase in $e_{33}$ is mainly due to the enhancement of $e_{33\_nonclamped}$. The values of $e_{33\_nonclamped}$ for $TM_{0.0625}Al_{0.9375}N$ (TM=Sc, Cr, Sr, Mo, Ru and Rh) are 10.0% higher than that of AlN. According to Equation 2, values of $z_{33}^*$ and $\frac{du}{d\delta}$ for the TM-alloy configurations are calculated and are provided in **Figures S4** and **S5** respectively. As shown in Figure S4, although the value of $z_{33}^*$ varies significantly among TMs, the average $z_{33}^*$ of all the cation ions are generally close to the value of Al in the pure AlN. This implies that the variation in $z_{33}^*$ is not the main contributor to the increase in $e_{33\_nonclamped}$. In contrast to $z_{33}^*$, the average value of $\frac{du}{d\delta}$ increases significantly in most of the TM-alloy configurations as shown in Figure S5. Values of $\frac{du}{d\delta}$ for $TM_{0.0625}Al_{0.9375}N$ (TM=Sc, Cr, Sr, Mo, Ru and Rh) are higher than 0.200, while the value for pure AlN is only 0.181. Taking $Ru_{0.0625}Al_{0.9375}N$ as an example, its $\frac{du}{d\delta}$ is 1.20 times that of AlN. The softening of $c_{33}$ in the TM-alloy configurations is shown in **Figure S6**. In particular, values of $c_{33}$ for TM=Mo, Ru and Rh, which are calculated to be 297, 310 and 308 GPa respectively, are much lower than that of AlN (377 GPa). With the lower $c_{33}$ and the larger $\frac{du}{d\delta}$, $TM_{0.0625}Al_{0.9375}N$ (TM=Sr, Mo, Ru and Rh) show higher $d_{33}$. It should be noted that the elastic constants of the TM-alloy AlN are still much higher than that of ZnO,[2] which implies that these configurations are suitable for high-frequency applications. To clearly show the contributions of $\frac{du}{d\delta}$ and $c_{33}$ to the increase in $d_{33}$, the relationships between $\frac{du}{d\delta}$ and $d_{33}$, and between $c_{33}$ and $d_{33}$ are provided in **Figure S7**. $d_{33}$ is generally proportional to $\frac{du}{d\delta}$ and inversely proportional to $c_{33}$. A better linear relationship between $d_{33}$ and $\frac{du}{d\delta}$ implies that the variation in $\frac{du}{d\delta}$ contributes more to the increase of $d_{33}$.

To further understand the variation in $\frac{du}{d\delta}$, the relaxed structures of the pure AlN, $TM_{0.0625}Al_{0.9375}N$ (TM=Zn and Rb) with lower $\frac{du}{d\delta}$ and $TM_{0.0625}Al_{0.9375}N$ (TM=Mo, Ru and Rh) with higher $\frac{du}{d\delta}$ are selected and are shown in **Figure 3**. Compared to the atom positions in AlN, the N atom above Zn and Zn (the N-Zn dipole) in $Zn_{0.0625}Al_{0.9375}N$ both move up slightly, and the N-Rb dipole in $Rb_{0.0625}Al_{0.9375}N$ moves up more significantly. In $Mo_{0.0625}Al_{0.9375}N$, the N atom above Mo moves up, but Mo moves down. In $TM_{0.0625}Al_{0.9375}N$ (TM=Ru and Rh), the N atoms above TM generally maintain their locations, as in pure AlN, and the TM atoms move down. The variation in $\frac{du}{d\delta}$ could be significantly correlated with local structural variations near the TM.

To explore the relationship between the structural change and the value of $\frac{du}{d\delta}$, pure AlN under uniaxial compressive strains along the $z$-axis is investigated, and the corresponding structural parameters and elastic and piezoelectric constants are provided in **Figure S8**. Al-N bonds in two different types denoted as $d_1$ (parallel to the $z$-axis) and $d_2$ (with an angle $\theta$ to the $z$-axis), show slight variations under compressive strain. The decrease in $c$ is mainly caused by the variation in $\theta$. In other words, $u$ and $cos\theta$ change significantly under compressive strain. Furthermore, $\frac{du}{d\delta}$ is analyzed to be proportional to $u$ and inversely proportional to $cos\theta$, according to **Equations S1-S4**. A more detailed analysis is provided in SI. As shown in Figures 3d-3f, the alloying of Ru and Rh increases $\theta$ between $d_1$ and $d_2$ in the TM-alloy sublayer, and the alloying of Mo increases $\theta$ in both the TM-alloy and its upper Al sublayers. Therefore, $\frac{du}{d\delta}$ increases in these three configurations. For the Zn- and Rb-alloy configurations, the TM decreases $\theta$ in the TM-alloy sublayer, but increases the angle in the upper Al sublayer. The opposite contributions of these two sublayers to $\frac{du}{d\delta}$ is further compared. A series spring model is proposed in **Figure S9**a. According to **Equations S5**, and **S6** and the corresponding discussion in SI, the changes in the thickness of the TM-alloy and its upper sublayers could be inversely proportional to the bond strength in these two sublayers. Therefore, the bond lengths and the minus integrated crystal orbital Hamilton populations (-ICOHP) of TM-N in both $d_1$ and $d_2$ are calculated and provided in **Figure S10**, and the corresponding atomic charges are given in **Figure S11**. From these two figures, TM-N generally shows a larger bond length and a weaker covalent bond strength than those of Al-N in pure AlN. The atomic charges of TM are also smaller than that of Al. Therefore, the spring coefficient of the TM-alloy sublayer is weaker than that of the upper Al sublayer, which implies that the thickness of the TM-alloy sublayer varies more under compressive strain. Therefore, the average values of $u$ and $\frac{du}{d\delta}$ are smaller in $TM_{0.0625}Al_{0.9375}N$ (TM=Zn and Rb). For a more convincing analysis, changes in the thickness of the Zn-alloy and its upper Al sublayers under uniaxial compressive strain are studied and the results are provided in Figure S9b. The change in sublayer thickness is in line with that of the series spring model, and a higher $\frac{du}{d\delta}$ is mainly due to the fact that the TM induces a weaker TM-N bond, and locates closer to the center of three neighboring N atoms. Additionally, $c_{33}$ depends on the bond strength of TM-N and the size of $cos\theta$ (as shown in Figure S8), and thus it couples to $\frac{du}{d\delta}$ and varies significantly in these TM-alloy configurations.

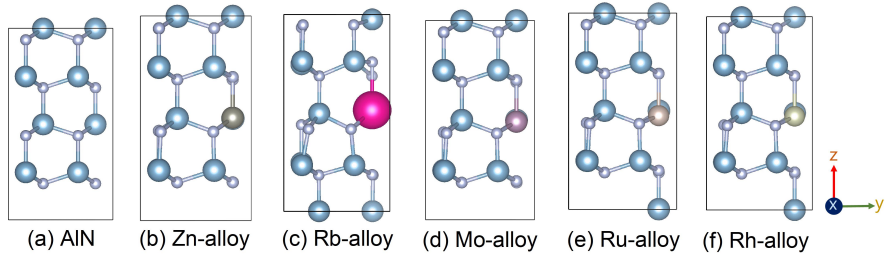

**Figure 3**. Relaxed structures of a) pure AlN; b)-f) Zn-, Rb-, Mo-, Ru- and Rh-alloy AlN respectively. The light blue and small grey balls denote Al and N respectively, and the balls in other colours denote TMs.

The increase in $\frac{du}{d\delta}$ and $d_{33}$ in TM-alloy AlN is explained by the local structural change near the TM, but what determines the location of the TM is not yet clear. As shown in Figure S5, the magnitude of $\frac{du}{d\delta}$ significantly depends on the group number of the TM. $TM_{0.0625}Al_{0.9375}N$ with TMs in group 1 (Na, K and Rb), group 7 (Mn and Tc), and groups 10-12 (Ni, Cu, Zn, Pd, Ag and Cd) generally show low values of $\frac{du}{d\delta}$, while configurations with TMs in group 6 (Cr and Mo) exhibit large values. A possible explanation is that the location of the TM is determined by the bonding behaviours near the TM, and the TM-N bond is correlated with the relative energies, types and occupation behaviour of the valence orbitals in the TM and N. The energies of valence orbitals in N, Al and TMs are provided in **Figure S12**. Al and the TMs possess valence electrons with their energies higher than or close to the N $2p$ orbital, and these electrons could contribute to the TM-N bonds. The $2p$ orbital of N, the $3p$ orbital of Al, and the $d$ orbitals in the TM are non-spherically dumbbell-like. If the $p$ and $d$ orbitals are not half-full or full, the valence electrons of the TM and Al could be more likely to bond with the N atoms below along the $z$-axis, because there are three N atoms at the bottom while only one N atom at the top of the TM. This conjecture is consistent with the fact that the bond length of Al-N in $d_1$ is longer than that in $d_2$ in pure AlN, as shown in Figure S10 a. With more valence electrons in the non-spherical orbitals of the TM, it could move down more significantly, such as in the Ru- and Rh-alloy AlN. Zn and Rb move up in $TM_{0.0625}Al_{0.9375}N$ (TM=Zn and Rb) because their contributed valence electrons are fewer and in the spherical $s$ orbital. When the $p$ and $d$ orbitals are half-full or full, the distribution of the valence electrons could approach a spherical shape, and the bond lengths of TM-N in the $d_1$ and $d_2$ types become closer. In Figure 3d, N moves up and Mo moves down, which could be ascribed to the larger atomic radius of Mo compared to Al. When the TM is in group 7, the spin behaviour of the valence electrons might also influence the location of the TM. With the three $2p$ electrons of N, a total of ten electrons could contribute to the TM-N bond when the TM is in group 7. As is well-known, the $d$ orbital could occupy ten electrons. Thus, the COHP of the Tc-N bond is spin-degenerate in $Tc_{0.0625}Al_{0.9375}N$, which is different from those of the Mo-N and Ru-N bonds in neighbouring $TM_{0.0625}Al_{0.9375}N$ (TM=Mo and Ru), as shown in **Figure S13**.

Once the influence of the TM type on $d_{33}$ is well-understood, the thermodynamic stability and piezoelectric performance of $TM_xAl_{1-x}N$ with varying $x$ could be further studied. Since $TM_{0.0625}Al_{0.9375}N$ (TM=Sr, Mo, Ru and Rh) show higher values of $d_{33}$ than those of the well-studied Sc- and Cr-alloy structures, variations in $x$ for $TM_xAl_{1-x}N$ (TM=Sc, Cr, Sr, Mo, Ru and Rh) are investigated. Regarding thermodynamic stability, the mixing enthalpies of $TM_xAl_{1-x}N$ as a function of $x$ are calculated as follows:

$$\Delta H(x) = [E_{TM_xAl_{1-x}N} - xE_{TMN} - (1-x)E_{AlN}]/N_{atom} \qquad (5)$$

where $E_{TM_xAl_{1-x}N}$, $E_{TMN}$ and $E_{AlN}$ are the total energies of $TM_xAl_{1-x}N$, TMN and AlN respectively, and $N_{atom}$ is the number of atoms in $TM_xAl_{1-x}N$. The stable AlN is shown in Figure 1, and the stable structures of TMN (TM=Sc, Cr, Sr, Mo, Ru and Rh) are selected as the most stable configurations listed on the OQMD website,[42] and their corresponding structures are provided in **Figure S14**. The relationships between $\Delta H(x)$ and $x$ in $TM_xAl_{1-x}N$ (TM=Sc, Cr, Sr, Mo, Ru and Rh) are provided in **Figure 4**a. At $x$=0.125, 0.1875 and 0.250, varying ranges of $\Delta H(x)$ are provided since fifteen SQSs are studied at each $x$. The horizontal grey dashed line shows $\Delta H(x)$ for the 0.200 $eV/atom$, which might be the possible threshold for judging the existence of these metastable configurations.[43] From the figure, $\Delta H(x)$ of the Sc- and Cr-alloy configurations are all lower than 0.200 $eV/atom$ up to $x$=0.250. $Sr_xAl_{1-x}N$ exhibits the highest $\Delta H(x)$ among the six TM-alloy configurations, and the values imply that $Sr_xAl_{1-x}N$ might be unstable when $x$ is higher than 0.100. For $TM_xAl_{1-x}N$ (TM=Mo, Ru and Rh), $\Delta H(x)$ generally decrease with the increasing group number of the TM, and the values increase with increasing $x$. This behaviour might be due to the difference in atomic radii between the TM and Al. All the SQSs of $Rh_{0.125}Al_{0.875}N$ are metastable. For $Rh_xAl_{1-x}N$ ($x$=0.1875 and 0.250) and $TM_xAl_{1-x}N$ (TM=Mo, Ru; $x$=0.125, 0.1875 and 0.250), only partial SQSs are metastable.

Figures 4b-4f show $d_{33}$ of $TM_xAl_{1-x}N$ for TM=Sc, Cr, Sr, Mo, Ru and Rh respectively. In Figures 4b and 4c, previous experimental results are also provided for comparison. The predicted average values are consistent with the experimental results. At a specific $x$, $d_{33}$ varies in different SQSs in all these TM-alloy configurations. The variation in $d_{33}$ has also been reported in previous theoretical and experimental work.[22, 25, 30, 44] In Figure 4, two different averaging approaches for $d_{33}$ are provided.

The blue hexagons show the average value of all the fifteen SQSs studied, while the cyan diamonds show the average value of the metastable SQSs. Since values of $\Delta H(x)$ for the Sr-alloy SQSs are all higher than 0.200 $eV/atom$, only the average values of the fifteen SQSs are provided in Figure 4d. Regarding the metastable $Sr_{0.0625}Al_{0.9375}N$, its $d_{33}$ is 1.49 times that of pure AlN. For the Mo-alloy structures, the average $d_{33}$ from all fifteen SQSs increases to 19.3 $pC/N$ at $x$=0.250. Excluding the unstable SQSs, the average value still increases to 13.6 $pC/N$ at $x$=0.1875. For $Ru_xAl_{1-x}N$, the average values of $d_{33}$ at $x$=0.125, 0.1875 and 0.250 from metastable SQSs are lower than 7.66 $pC/N$ at $x$=0.0625. The alloying of Rh also significantly enhances $d_{33}$, and the average $d_{33}$ from the metastable SQSs increases to 15.1 $pC/N$ at $x$=0.125.

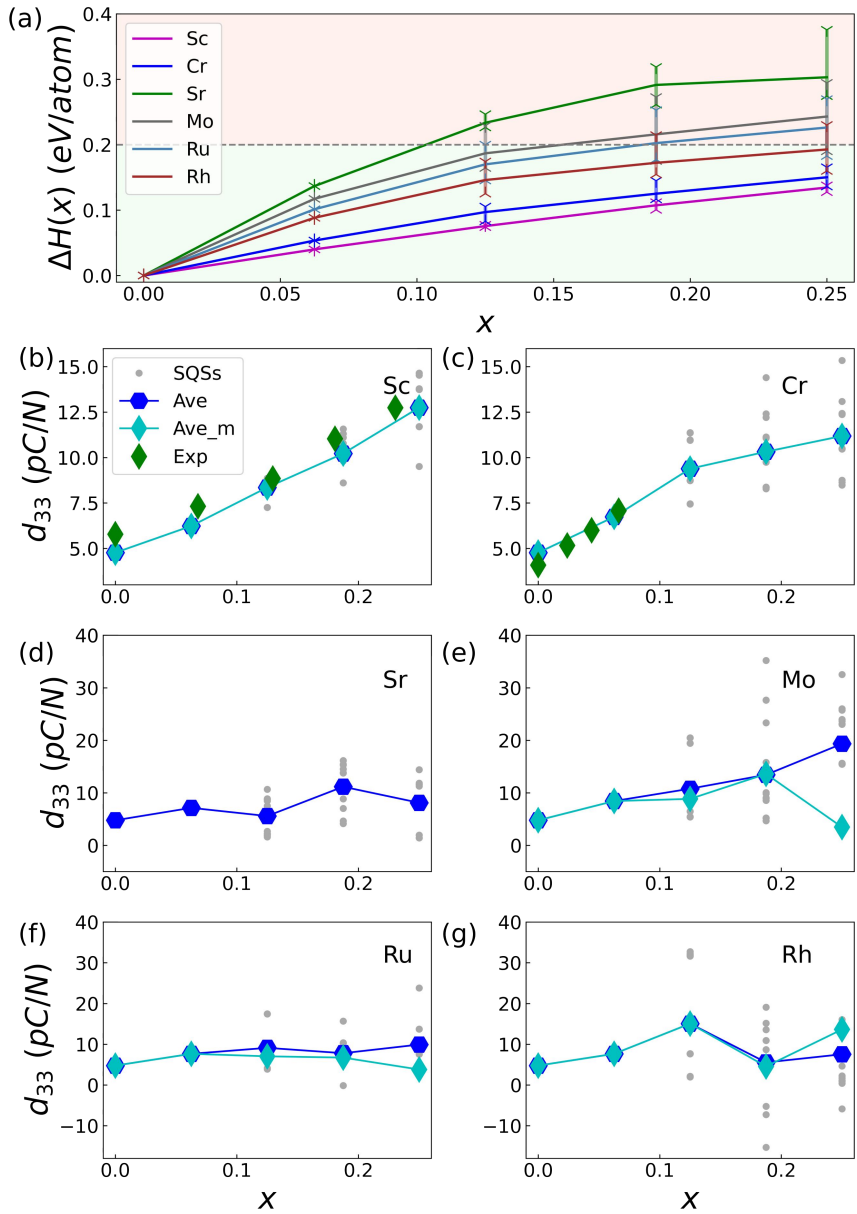

**Figure 4.** a) $\Delta H(x)$ of $TM_xAl_{1-x}N$ (TM=Sc, Cr, Sr, Mo, Ru and Rh). The magenta, blue, green, dark grey, steel blue and brown lines show the values of $TM_xAl_{1-x}N$ where the TMs are Sc, Cr, Sr, Mo, Ru and Rh, respectively; b)-g) $d_{33}$ of $TM_xAl_{1-x}N$ for TM=Sc, Cr, Sr, Mo, Ru and Rh respectively, where the grey dot shows the $d_{33}$ for each SQS and the blue hexagon represents the average $d_{33}$ of the fifteen SQSs studied, the cyan diamond shows the average value of the metastable SQSs, and the green diamonds in b) and c) show previous experimental values.[12, 13]

Both Mo- and Rh-alloying could significantly enhance the $d_{33}$ of AlN, but Rh-alloying is not suitable for large-scale applications because Rh is even more expensive than Sc. Therefore, only the Mo-alloy AlN is studied in further detail. Since the generalized gradient approximation (GGA) over-delocalizes the electron density, Hubbard U correction (GGA+U) is performed for the 4$d$ orbital of Mo.[45] With the same SQSs, values of $\Delta H(x)$ and $d_{33}$ for $Mo_xAl_{1-x}N$ with $x$ up to 0.250 from GGA and GGA+U are compared in **Figures 5**a and 5b respectively. Values of $\Delta H(x)$ from GGA+U are smaller than those from GGA. From GGA, five SQSs of $Mo_{0.125}Mo_{0.875}N$, eight SQSs of $Mo_{0.1875}Al_{0.8125}N$ and fourteen SQSs of $Mo_{0.250}Al_{0.750}N$ show $\Delta H(x)$ higher than 0.200 $eV/atom$. However, only one SQS of $Mo_{0.250}Al_{0.750}N$ is unstable from GGA+U. In Figure 5b, values of $d_{33}$ for all of the SQSs and the average values from metastable SQSs are provided. Compared to those of GGA, the variation range of $d_{33}$ at each $x$ from GGA+U is also slightly smaller, and the average values increase up to $x$=0.250. More specifically, the average $d_{33}$ are 4.77, 8.05, 12.3, 12.7 and 17.2 $pC/N$ at $x$=0.00, 0.0625, 0.125, 0.1875 and 0.250 respectively. The markedly different average $d_{33}$ values at $x$=0.250 from GGA and GGA+U could be ascribed to the fact that the value from GGA is from just one metastable SQS, and that from GGA+U is averaged from fourteen metastable SQSs.

All of the above configurations are based on the $2\times2\times2$ AlN supercell. The influence of supercell size is then further studied. As shown in Figures 5c and 5d, $\Delta H(x)$ and $d_{33}$ of $Mo_xAl_{1-x}N$ based on the 2×2×2, 3×3×1 and 2×2×3 AlN supercells are compared on basis of GGA+U. In Figure 5c, $\Delta H(x)$ increases with increasing $x$ in all supercells. At a specific $x$, values of $\Delta H(x)$ of the 2×2×2 and 2×2×3 supercells are similar, while the values of the 3×3×1 supercells are a little smaller. Apart from one SQS of $Mo_{0.250}Al_{0.750}N$ from the 2×2×2 supercells, all of the Mo-alloy configurations are metastable. Based on these metastable structures, the average values of $d_{33}$ is provided in Figure 5d. Compared to the values from the 2×2×2 supercells, the average values of $d_{33}$ for the 3×3×1 supercells are a little lower, which are 7.45, 9.10, 11.4 and 11.2 $pC/N$ at $x$=0.0556, 0.111, 0.167 and 0.222 respectively. The values for the 2×2×3 supercells are generally between those of the 3×3×1 and 2×2×3 supercells, except for a high value of 17.2 $pC/N$ determined at $x$=0.167. It is worth noting that a large value of 58.6 $pC/N$ is found for a metastable SQS at this alloy content, which is 12.3 times that of the pure AlN. On the basis of the piezoelectric performance of these different supercells, it can be concluded that Mo-alloying can significantly improve the $d_{33}$ of AlN. Moreover, the varying values for SQSs imply that $d_{33}$ is dependent on the coordinates of Mo.

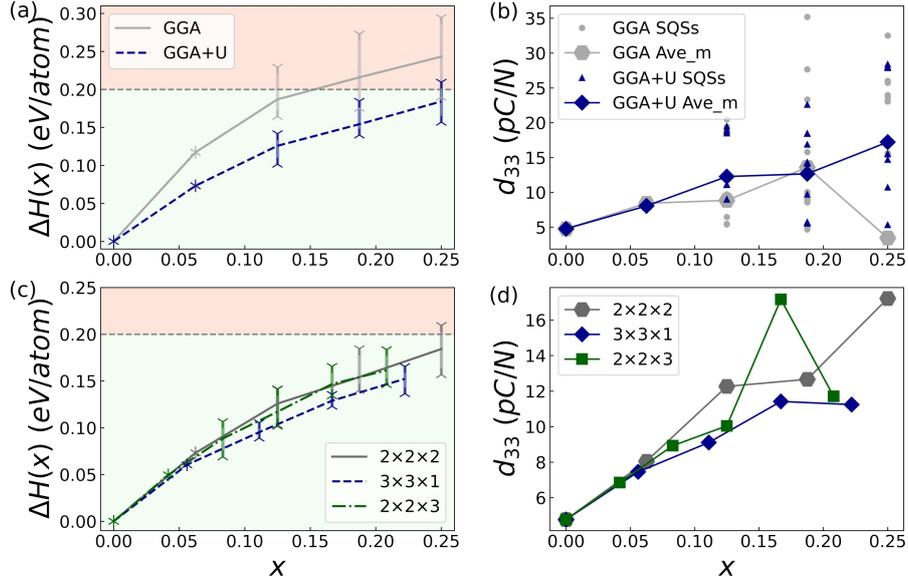

**Figure 5.** Values of a) *ΔH(x)* and b) *d₃₃* of Mo$_x$Al$_{1-x}$N based on the 2×2×2 AlN supercell, where the grey and blue lines show the values from GGA and GGA+U respectively, and the grey dashed line in a) shows the value of 0.200 *eV/atom*, while the grey dots, and blue triangles in b) represent the values of SQSs, and the grey hexagons, and blue diamonds show the average *d₃₃* of the metastable SQSs from GGA and GGA+U respectively; c) *ΔH(x)* and d) average *d₃₃* of the metastable SQS of Mo$_x$Al$_{1-x}$N from GGA+U, where the grey, blue and green lines and symbols show the corresponding values for the 2×2×2, 3×3×1 and 2×2×3 supercells respectively.

Regarding the influence of Mo coordinates on the value of $d_{33}$ for Mo$_x$Al$_{1-x}$N, the average Mo-Mo distance, $r_{ave}$, and the ratio of the number of the Mo-alloy sublayers to that of the total Al sublayers along the *z*-axis, $R_N$, are found to be more strongly correlated. $r_{ave}$ is calculated as follows:

$$r_{ave} = \frac{r_i g(r_i)}{\sum_{i=0}^{r_{max}} g(r_i)} \tag{6}$$

where $g(r_i)$ denotes the radial distribution function of the Mo atoms. $R_N$ is obtained directly from the Mo-alloy configuration. For the Mo$_{0.0625}$Al$_{0.9375}$N shown in Figure 3d as an example, there are a total of four Al sublayers along the *z*-axis, only one of which contains an Mo atom. Thus, $R_N$ is equal to 0.250 in this configuration. The relationships between $r_{ave}$ and $d_{33}$ and between $R_N$ and $d_{33}$ for all of the Mo-alloy SQSs from GGA+U are provided in **Figures 6**a and 6b respectively. In Figure 6a, $r_{ave}$ generally shows a larger range when *x* is lower in Mo$_x$Al$_{1-x}$N. With the same *x*, a higher $d_{33}$ is more likely to be obtained when $r_{ave}$ is larger. From Figure 6b, it is possible to obtain a larger $d_{33}$ when $R_N$ is larger. For the 2×2×2 supercells, the largest $R_N$ is 0.750, and the largest $d_{33}$ of 28.4 *pC/N* occurs at this $R_N$ in a SQS of Mo$_{0.250}$Al$_{0.750}$N. The largest $R_N$ of the 3×3×1 supercells is 1.00, and the largest $d_{33}$ of 19.2 *pC/N* is obtained at this $R_N$ in a SQS of Mo$_{0.222}$Al$_{0.778}$N. Similarly, the largest $R_N$ is 0.667 in the 2×2×3 supercells, and the largest $d_{33}$ of 58.6 $pC/N$ is found at this $R_N$ in a SQS of $Mo_{0.167}Al_{0.833}N$. From Figure 6, it could be speculated that it is possible to obtain a higher $d_{33}$ in an $Mo_xAl_{1-x}N$ configuration with more Al sublayers along the z-axis containing Mo atoms (for a larger $R_N$), and with a greater difference in the $xy$ coordinates of the Mo atoms between neighbouring sublayers (for a larger $r_{ave}$). For a more convincing analysis, four SQSs of $Mo_{0.250}Al_{0.750}N$ with markedly different $d_{33}$ values from the 2×2×2 supercells are provided in **Figure S15**. The same number of SQSs with different $d_{33}$ of $Mo_{0.222}Al_{0.778}N$ from the 3×3×1 supercells, and those of $Mo_{0.167}Al_{0.833}N$ from the 2×2×3 supercells, are provided in **Figures S16** and **S17** respectively. The value of $d_{33}$ in these different SQSs is generally in line with the deduction above. Taking the SQSs in Figure S15 as an example, values of $R_N$ in Figures S15a and S15b are both 0.750, but the $xy$ coordinates of the Mo atoms between neighbouring sublayers are more different in the latter. For SQSs in Figures S15c and S15d, their Mo atoms are both scattered in the $xy$ plane, but their values of $R_N$ are different. The value of $R_N$ in Figure S15c is 0.500, while in Figure S15d is 0.750. As a consequence, $d_{33}$ in Figure S15b is larger than that in Figure 15a, and the value in Figure S15d is larger than that in Figure S15c. The value of $d_{33}$ in Figure S15d is the largest of all.

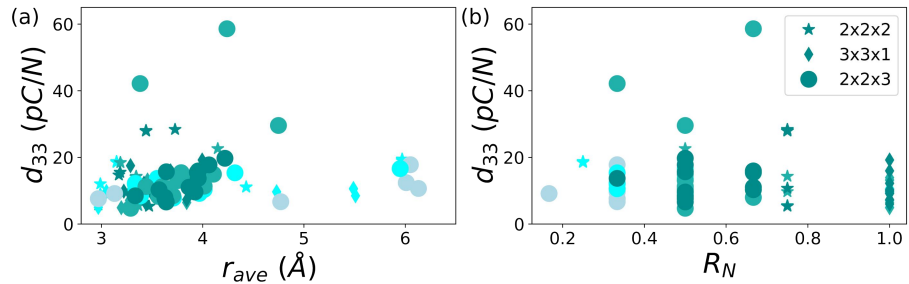

**Figure 6.** Relationships between $d_{33}$ and: a) $r_{ave}$; b) $R_N$. The star, diamond and circle symbols denote data from the 2×2×2, 3×3×1 and 2×2×3 supercells respectively. Different colours represent the varying $x$ in $Mo_xAl_{1-x}N$, with the symbols in light blue, cyan, light sea green, and dark cyan representing values of $x$ in the ranges of $x\leq0.100$, $0.100<x\leq0.150$, $0.150<x\leq0.200$, and $0.200<x\leq0.250$ respectively.

The underlying mechanism of the influence of Mo coordinates on $d_{33}$ is subsequently further studied. As discussed above, the increase in $d_{33}$ is due to the fact that Mo moves down and the N above it moves up, which introduces a larger average $u$ and weaker Mo-N bonds. As shown in Figures S15-S17, the SQSs with lower $d_{33}$ generally have a common feature that they have dimer or even a line of Mo atoms along the z-axis. The Mo atoms in the dimer generally become closer to each other, which counteracts the behaviour of the Mo atom in moving down. The two Mo atoms in Figure S17a even move up compared to the neighbouring Al atoms. As a result, levels of $d_{33}$ in these configurations are generally low because the mechanism for improving $d_{33}$ is weakened by the Mo dimer. The Mo atoms in dimers may get closer because the $4d$ orbital of Mo is not in a stable electronic configuration (empty or full) after bonding with neighbouring N atoms. Then, the Mo atoms in a dimer become closer to bond and share their electrons. This behaviour might be verified by the COHP between the Mo atoms in the dimer. **Figure S18** shows the COHP of the Mo-Mo dimer in the configuration shown in Figure S15a. The COHP below and above the Fermi level are both in bonding states, which implies that the Mo atoms in the dimer are prone to become closer and to share their electrons. On the other hand, in the configuration with more sublayers along the $z$-axis containing Mo atoms, $d_{33}$ could increase more significantly since each Mo-alloy sublayer can contribute to the increase in $\dfrac{du}{d\delta}$ and the softening of $c_{33}$. Therefore, to obtain a higher $d_{33}$ in the Mo-alloy AlN, it is suggested to alloy the Mo atoms in more sublayers and to avoid the formation of dimers of Mo atoms along the $z$-axis.

## 3. Conclusion

In summary, we have studied for the first time the influence on $d_{33}$ of the type and the distribution of TMs in alloys of AlN. Regarding the influence of TM type, the value of $d_{33}$ for a TM-alloy AlN is determined to be significantly correlated with the group number of the TM. Alloys of $TM_{0.0625}Al_{0.9375}N$ with a TM in groups 1, 7, 10-12 generally show lower $d_{33}$, while configurations with a TM in group 6 present larger values. Among the twenty-eight $TM_{0.0625}Al_{0.9375}N$ configurations studied, most show higher $d_{33}$ than that of pure AlN. $TM_{0.0625}Al_{0.9375}N$ (TM=Sr, Mo, Ru and Rh) even show higher values than the well-studied Sc- and Cr-alloys of AlN at the same TM content. The increase in $d_{33}$ is ascribed to the fact that the TM moves down to the centre of three neighbouring N atoms below, which leads to a larger average $u$ and weaker TM-N bonds. The thermodynamic stability and piezoelectric constants of $TM_xAl_{1-x}N$ (TM=Sc, Cr, Sr, Mo, Ru and Rh) with varying values of $x$ up to 0.250 are further studied. The Mo alloy is more effective in improving $d_{33}$ on the basis of piezoelectric performance and the cost of the TM. The influence of Hubbard U correction and supercell size on the performance of the Mo-alloy configurations is also studied. From different SQSs, an average $d_{33}$ of 17.2 $pC/N$ is determined for $Mo_{0.250}Al_{0.750}N$. A high value of 12.3 times that of pure AlN is realized in a metastable SQS of $Mo_{0.167}Al_{0.833}N$. The value of $d_{33}$ is dependent on the coordinates of Mo in $Mo_xAl_{1-x}N$. A higher value is prone to appear in a configuration with more Al sublayers along the $z$-axis containing Mo atoms, and the $xy$ coordinates of the Mo atoms differ more between neighbouring sublayers. The latter requirement avoids the formation of Mo-Mo dimers along the $z$-axis, because this dimer can significantly weaken the mechanism for improving $d_{33}$. Generally, this work provides instructive guidance for experimental scientists aiming to enhance the $d_{33}$ of AlN through TM alloying. The underlying mechanisms of the influence of TM type and distribution on $d_{33}$ have been revealed. Moreover, Mo-alloying significantly improves the $d_{33}$ of AlN, and we look forward to the successful synthesis of Mo-alloys of AlN in the near future.

## 4. Computational details

$TM_{0.0675}Al_{0.9325}N$ was constructed by replacing an Al atom with a TM on the basis of the $2\times2\times2$ AlN supercell (32 atoms). In investigating varying alloying contents in $TM_xAl_{1-x}N$, values of $x$ up to 0.250 increasing in steps of 0.0625 were studied on basis of the $2\times2\times2$ AlN supercells. For the Mo-alloy configurations, $2\times2\times3$ (48 atoms) and $3\times3\times1$ (36 atoms) AlN supercells were also studied. For the $2\times2\times3$ supercells, values of $x$ up to 0.208 with a step size of 0.0417, and for the $3\times3\times1$ supercells, values of $x$ up to 0.222 with a step size of 0.0556, were investigated. At each values of $x$, fifteen SQSs[46, 47] were constructed when the number of TM atoms in the supercell was greater than or equal to two.

First-principles calculations were implemented in the Vienna ab initio simulation package (VASP).[48, 49] The plane wave basis size with projector augmented wave (PAW) pseudopotential[50] was employed. GGA of Perdew-Burke-Ernzerhof (GGA-PBE)[51] was adopted to describe the exchange and correlation functional, and the plane-wave cutoff energy was set as 540 eV. All structures were relaxed until the forces on each atom were less than $1.0\times10^{-4}$ $eV/Å$. For the $2\times2\times2$ AlN supercell, a $\Gamma$-centered $6\times6\times4$ k-point mesh was adopted, while $4\times4\times6$ and $6\times6\times2$ k-point meshes were employed for the $3\times3\times1$ and $2\times2\times3$ AlN supercells, respectively. The piezoelectric and elastic constants were calculated on the basis of density functional perturbation theory (DFPT).[52, 53] For the Mo-alloy configurations, both GGA and GGA+U functionals were studied. An effective Hubbard U parameter of 3.83 $eV$ was adopted for the $4d$ orbital of Mo.[45, 54] To view the structural change under uniaxial strains, the unit cells of the pure AlN and the Zn-alloy AlN under compressive strain along the $z$-axis were investigated. This simulation was implemented by recompiling the *constr_cell_relax.F* script in the VASP code. The atomic charge was investigated using Bader charge analysis[55] on basis of a $180\times180\times180$ grid mesh. Bond strength was calculated according to -ICOHP using a combination of VASP and LOBSTER[56] codes. All structures were visualized in VESTA3 code.[57]

## Acknowledgements

The authors acknowledge financial support from the National Natural Sciences Foundation of China (Grant Nos. 11974252, 11604346, 21603146), the Research and Development Program in the Significant Area of Guangdong Province (Grant No. 2020B0101040002), the special Project in Key Fields of Colleges in Guangdong Province (2020ZDZX2007), the Research Project in Fundamental and Application Fields of Guangdong Province (2020A1515110561), and the Shenzhen Science and Technology Project (Grant Nos. SGDX20190919102801693, JCYJ20180507182106754, JCYJ20180507182439574, RCBS20200714114918249, GJHZ20200731095803010).

## Conflict of Interest
The authors declare no conflict of interest.

## Author Contributions
X. -H. Z. and X. M. contributed equally to this work.

**References:**

1. P. Muralt, *J. Am. Ceram. Soc.* **2008**, *91,* 1385.
2. Y.Q. Fu, J.K. Luo, N.T. Nguyen, A.J. Flewitt, X.T. Zu, Y. Li, G. McHale, A. Matthews, E. Iborra, H. Du, W.I. Milne, *Prog. Mater. Sci.* **2017**, *89,* 31.
3. H.P. Loebl, M. Klee, C. Metzmacher, W. Brand, R. Milsom, P. Lok, *Mater. Chem. Phys.* **2003**, *79,* 143.
4. G. Piazza, V. Felmetsger, P. Muralt, R. H. Olsson III, R. Ruby, *MRS Bull.* **2012**, *37,* 1051.
5. A. Wixforth, *JALA,* **2016**, *11,* 399.
6. R.K. Cook, P.G. Weissler, *Phys. Rev.* **1950**, *80,* 712.
7. R. Bechmann, *Phys. Rev.* **1958**, *110,* 1060.
8. R. Weigel, D.P. Morgan, J.M. Owens, A. Ballato, K.M. Lakin, K. Hashimoto, C.C.W. Ruppel, *IEEE Trans. Microwave Theory Tech.* **2002**, *50,* 738.
9. U.C. Kaletta, C. Wipf, M. Fraschke, D. Wolansky, M.A. Schubert, T. Schroeder, C. Wenger, *IEEE Trans. Electron Devices* **2015**, *62,* 764.
10. C. Caliendo, P. Imperatori, *Appl. Phys. Lett.* **2003**, *83,* 1641.
11. T. Aubert, J. Bardong, O. Legrani, O. Elmazria, M. B. Assouar, G. Bruckner, A. Talbi, *J. Appl. Phys.* **2013**, *114,* 014505.
12. S. Manna, K.R. Talley, P. Gorai, J. Mangum, A. Zakutayev, G. L. Brennecka, V. Stevanović, C.V. Ciobanu, *Phys. Rev. Appl.* **2018**, *9,* 034026.
13. M. Akiyama, T. Kamohara, K. Kano, A. Teshigahara, Y. Takeuchi, N. Kawahara, *Adv. Mater.* **2009**, *21,* 593.
14. M. Akiyama, K. Kano, A. Teshigahara, *Appl. Phys. Lett.* **2009**, *95,* 162107.
15. M. Akiyama, K. Umeda, A. Honda, T. Nagase, *Appl. Phys. Lett.* **2013**, *102,* 021915.
16. F. Tasnádi, B. Alling, C. Höglund, G. Wingqvist, J. Birch, L. Hultman, I. A. Abrikosov, *Phys. Rev. Lett.* **2010**, *104,* 137601.
17. W. Wang, P. M. Mayrhofer, X. He, M. Gillinger, Z. Ye, X. Wang, A. Bittner, U. Schmid, J.K. Luo, *Appl. Phys. Lett.* **2014**, *105,* 133502.
18. R. Matloub, M. Hadad, A. Mazzalai, N. Chidambaram, G. Moulard, C. S. Sandu, Th. Metzger, P. Muralt, *Appl. Phys. Lett.* **2013**, *102,* 152903.
19. T. Yanagitani, M. Suzuki, *Appl. Phys. Lett.* **2014**, *105,* 122907.
20. Q. Zhang, T. Han, J. Chen, W. Wang, K. Hashimoto, *Diamond Relat. Mater.* **2015**, *58,* 31.
21. M. Moreira, J. Bjurström, I. Katardjev, V. Yantchev, *Vacuum* **2011**, *86,* 23.
22. F. Tasnádi, I.A. Abrikosov, I. Katardjiev, *Appl. Phys. Lett.* **2009**, *94,* 151911.
23. L. Liljeholm, M. Junaid, T. Kubart, J. Birch, L. Hultman, I. Katardjiev, *Surf. Coat. Technol.* **2011**, *206,* 1033.
24. A. Žukauskaitė, C. Tholander, J. Palisaitis, P.O.A. Persson, V. Darakchieva, N.B. Sedrine, F. Tasnádi, B. Alling, J. Birch, L. Hultman, *J. Phys. D: Appl. Phys.* **2012**, *45,* 422001.

25. P.M. Mayrhofer, H. Riedl, H. Euchner, M. Stöger-Pollach, P.H. Mayrhofer, A. Bittner, U. Schmid, *Acta Mater.* **2015**, *100,* 81.

26. S. Manna, G.L. Brennecka, V. Stevanović, C.V. Ciobanu, *J. Appl. Phys.* **2017**, *122,* 105101.

27. X. Hu, Z. Tai, C. Yang, *Mater. Lett.* **2018**, *217,* 281.

28. H. Liu, F. Zeng, G. Tang, F. Pan, *Appl. Surf. Sci.* **2013**, *270,* 225.

29. T. Yanagitani, J. Jia. *IEEE Int. Ultrason. Symp.* **2019**, 894.

30. M. Uehara, Y. Amano, S.A. Anggraini, H. Yamada, M. Akiyama, *Ceram. Int.* **2021**, *47,* 16029.

31. J.T. Luo, B. Fan, F. Zeng, F. Pan, *J. Phys. D: Appl. Phys.* **2009**, *42,* 235406.

32. F. Laidoudi, S. Amara, C. Caliendo, F. Boubenider, F. Kanouni, A. Assali, *Appl. Phys. A: Mater. Sci. Process.* **2021**, *127,* 255.

33. K. Hirata, H. Yamada, M. Uehara, S.A. Anggraini, M. Akiyama, *ACS Omega* **2019**, *4,* 15081.

34. K. Hirata, H. Yamada, M. Uehara, S.A. Anggraini, M. Akiyama, *J. Phys. Chem. Solids* **2021**, *152,* 109913.

35. M. Uehara, H. Shigemoto, Y. Fujio, T. Nagase, Y. Aida, K. Umeda, M. Akiyama, *Appl. Phys. Lett.* **2017**, *111,* 112901.

36. T. Yokoyama, Y. Iwazaki, T. Nishihara, J. Tsutsumi, *IEEE Int. Ultrason. Symp.* **2016**.

37. T. Yokoyama, Y. Iwazaki, Y. Onda, Y. Sarajjima, T. Nishihara, M. Ueda, *IEEE Int. Ultrason. Symp.* **2014**, 281.

38. M. Noor-A-Alam, O. Z. Olszewski, H. Campanella, M. Nolan. *ACS Appl. Mater. Interfaces* **2021**, *13,* 944.

39. X.-H. Zha, X. Ma, S. Du, R.-Q. Zhang, R. Tao, J.-T. Luo, C. Fu, *Inorg. Chem.* **2022**, *61,* 2129.

40. F. Bernardini, V. Fiorentini, D. Vanderbilt, *Phys. Rev. B* **1997**, *56,* R10024.

41. K. Hirata, Y. Mori, H. Yamada, M. Uehara, S.A. Anggraini, M. Akiyama, *Materials*, **2021**, *14,* 309.

42. S. Kirklin, J.E. Saal, B. Meredig, A. Thompson, J.W. Doak, M. Aykol, S. Rühl, C. Wolverton, *npj Comput. Mater.* **2015**, *1,* 15010.

43. S.P. Ong, L. Wang, B. Kang, G. Ceder, *Chem. Mater.* **2008**, *20,* 1798.

44. E. Wistrela, I. Schmied, M. Schneider, M. Gillinger, P.M. Mayrhofer, A. Bittner, U. Schmid, *Thin Solid Films* **2018**, *648,* 76.

45. J.W. Bennett, B.G. Hudson, I.K. Metz, D. Liang, S. Spurgeon, Q. Cui, S.E. Mason, *Comput. Mater. Sci.* **2019**, *170,* 109137.

46. A. Zunger, S.-H. Wei, L.G. Ferreira, J.E. Bernard, *Phys. Rev. Lett.* **1990**, *65,* 353.

47. A. van de Walle, P. Tiwary, M. de Jong, D.L. Olmsted, M. Asta, A. Dick, D. Shin, Y. Wang, L.-Q. Chen, Z.-K. Liu, *CALPHAD: Comput. Coupling Phase Diagrams Thermochem.* **2013**, *42,* 13.

48. G. Kresse, J. Furthmüller, *Comput. Mater. Sci.* **1996**, *6,* 15.

49. G. Kresse, J. Furthmüller, *Phys. Rev. B* **1996**, *54,* 11169.

50. G. Kresse, D. Joubert, *Phys. Rev. B* **1999**, *59,* 1758.

51. J.P. Perdew, K. Burke, M. Ernzerhof, *Phys. Rev. Lett.* **1996**, *77,* 3865.

52. X. Gonze, C. Lee, *Phys. Rev. B* **1997**, *55,* 10355.

53. X. Wu, D. Vanderbilt, D.R. Hamann, *Phys. Rev. B* **2005**, *72,* 035105.

54. M. Cococcioni, S. de Gironcoli, *Phys. Rev. B* **2005**, *71,* 035105.

55. W. Tang, E. Sanville, G. Henkelman, *J. Phys.: Condens. Matter* **2009**, *21,* 084204.

56. R. Nelson, C. Ertural, J. George, V.L. Deringer, G. Hautier, R. Dronskowski, *J. Comput. Chem.* **2020**, *41*, 1931.

57. K. Momma, F. Izumi, *J. Appl. Crystallogr.* **2011**, *44,* 1272.

**ToC figure**

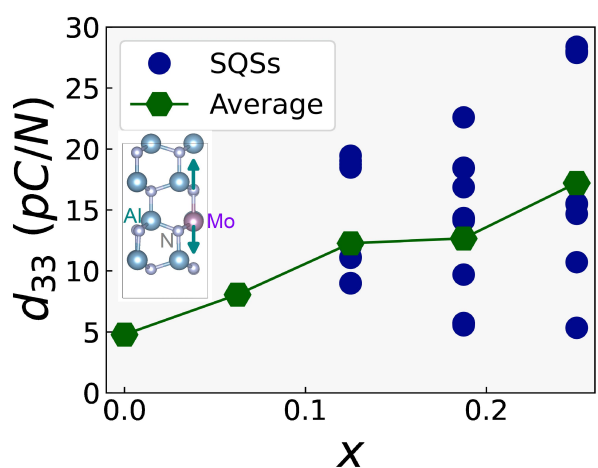

Supporting Information

# Enhanced piezoelectric response of AlN via alloying of transitional metals, and influence of type and distribution of transition metals

*Xian-Hu Zha[+], Xiufang Ma[+], Jing-Ting Luo, Chen Fu[*]*

X.-H. Zha, X. Ma, J.-T. Luo, C. Fu
Key Laboratory of Optoelectronic Devices and Systems of Ministry of Education and Guangdong Province, College of Physics and Optoelectronic Engineering, Shenzhen University, Shenzhen 518060, China. [*]Email: chenfu@szu.edu.cn.
[+]These authors contributed equally to this work.

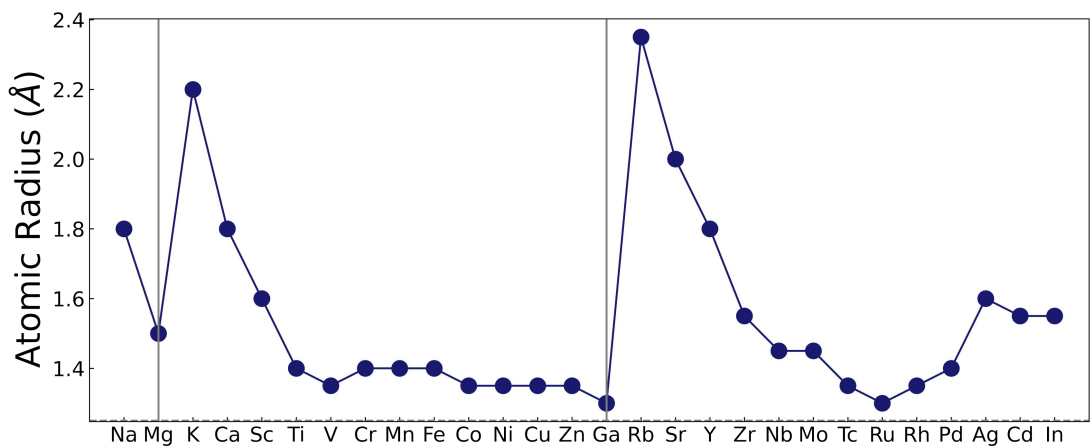

**Figure S1**. Atomic radii of TMs considered in this work. The bottom dashed line represents the atomic radius of Al.

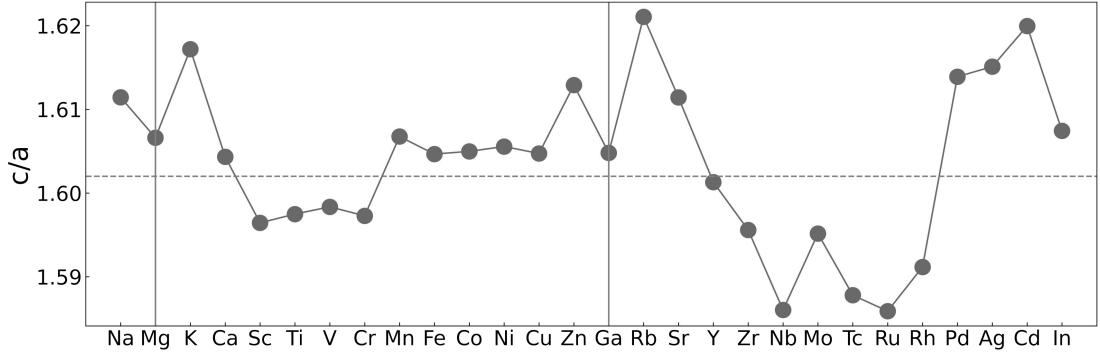

**Figure S2.** Ratio of $c/a$ in $TM_{0.0625}Al_{0.9375}N$. The grey dashed line denotes the corresponding value in pure AlN.

**Table S1.** Piezoelectric stress constants $d_{ij}$ (i=1-3, j=1-6) of $TM_{0.0625}Al_{0.9375}N$.

| TM | $d_{15}$ | $d_{16}$ | $d_{21}$ | $d_{22}$ | $d_{23}$ | $d_{24}$ | $d_{31}$ | $d_{32}$ | $d_{33}$ |
|---|---|---|---|---|---|---|---|---|---|
| Na | -5.72 | -0.286 | -0.142 | 0.142 | 0.000 | -5.71 | -2.09 | -2.09 | **4.82** |
| Mg | -6.73 | -5.70 | -2.86 | 2.89 | -0.0315 | -6.72 | -2.32 | -2.32 | **5.48** |
| K | -3.91 | -0.713 | -0.356 | 0.355 | 0.001 | -3.89 | -2.45 | -2.45 | **5.49** |
| Ca | -4.89 | 5.61 | 2.79 | -2.79 | 0.000 | -4.87 | -2.63 | -2.63 | **6.28** |
| Sc | -3.08 | 0.635 | 0.316 | -0.318 | 0.001 | -3.06 | -2.61 | -2.61 | **6.24** |
| Ti | -7.30 | 4.76 | 2.42 | -2.42 | -0.001 | -7.34 | -2.35 | -2.35 | **6.03** |
| V | -2.62 | -0.009 | -0.004 | -0.004 | 0.000 | -2.58 | -2.34 | -2.34 | **5.93** |
| Cr | -8.53 | 2.02 | 1.02 | -1.04 | 0.0170 | -9.93 | -2.68 | -2.68 | **6.74** |
| Mn | -37.1 | 10.4 | 5.28 | -5.10 | -0.201 | -36.9 | -2.09 | -2.09 | **4.98** |
| Fe | -2.72 | 0.018 | 0.009 | -0.009 | 0.000 | -2.70 | -2.13 | -2.13 | **5.06** |
| Co | -3.75 | 1.33 | 0.669 | -0.668 | -0.001 | -3.73 | -2.21 | -2.20 | **5.33** |
| Ni | -2.59 | 0.335 | 0.168 | -0.168 | 0.000 | -2.57 | -2.06 | -2.06 | **4.87** |
| Cu | -12.0 | -0.569 | -0.275 | 0.299 | -0.027 | -12.0 | -2.11 | -2.11 | **5.18** |
| Zn | -16.7 | -0.818 | -0.306 | 0.771 | -0.513 | -16.9 | 0.473 | 0.473 | **-0.615** |
| Ga | -3.21 | -0.071 | -0.035 | 0.035 | 0.000 | -3.20 | -2.04 | -2.04 | **4.84** |
| Rb | -3.16 | -0.583 | -0.290 | 0.290 | 0.000 | -3.14 | -2.38 | -2.38 | **4.74** |
| Sr | -5.37 | -5.01 | -2.52 | 2.51 | 0.008 | -5.35 | -3.18 | -3.18 | **7.13** |
| Y | -3.85 | 1.31 | 0.656 | -0.656 | 0.000 | -3.84 | -2.49 | -2.49 | **5.91** |
| Zr | -28.8 | 13.8 | 7.54 | -7.56 | -0.050 | -30.2 | -2.25 | -2.25 | **5.71** |
| Nb | -2.92 | 0.609 | 0.305 | -0.305 | 0.000 | -2.90 | -2.09 | -2.09 | **5.64** |
| Mo | -8.58 | 11.6 | 4.29 | -7.31 | 3.67 | -7.49 | -3.28 | -3.28 | **8.43** |
| Tc | -3.68 | 0.690 | 0.345 | -0.345 | 0.000 | -3.67 | -2.00 | -2.00 | **5.11** |
| Ru | -5.06 | 6.02 | 3.00 | -3.00 | 0.000 | -5.06 | -2.66 | -2.66 | **7.66** |
| Rh | -15.6 | 6.77 | 3.45 | -3.30 | -0.207 | -15.4 | -3.16 | -3.16 | **8.32** |
| Pd | -5.84 | -3.39 | -1.70 | 1.70 | 0.001 | -5.82 | -2.28 | -2.28 | **5.27** |
| Ag | -11.9 | -0.293 | -0.146 | 0.152 | -0.005 | -11.9 | -2.02 | -2.02 | **4.64** |
| Cd | -4.71 | -1.60 | -0.802 | 0.805 | -0.003 | -4.70 | -2.04 | -2.04 | **4.58** |
| In | -2.84 | -0.867 | -0.436 | 0.436 | 0.000 | -2.82 | -2.09 | -2.09 | **4.88** |

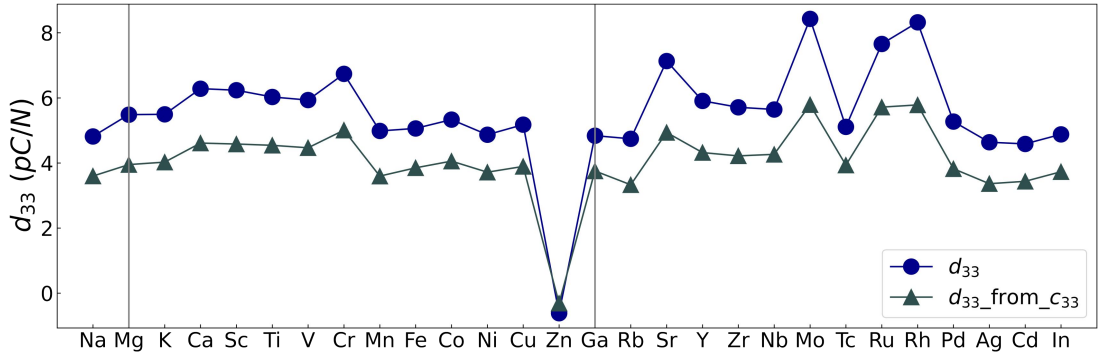

**Figure S3**. $d_{33}$ of $TM_{0.0625}Al_{0.9375}N$ from two different calculation methods.

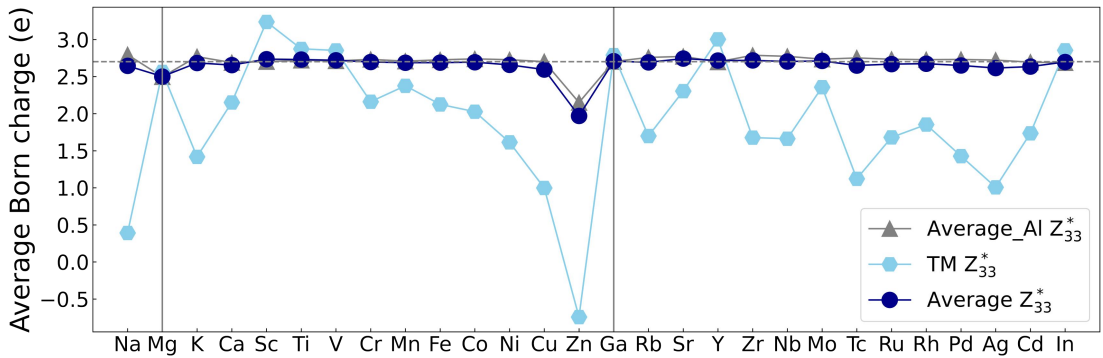

**Figure S4.** $Z_{33}^*$ of $TM_{0.0625}Al_{0.9375}N$. The grey triangles represent the average value of Al ions. The light blue hexagons show the value of TM. The dark blue circles represent the average value of all the Al and TM cation ions. The dashed grey line denotes the corresponding value of Al in pure AlN.

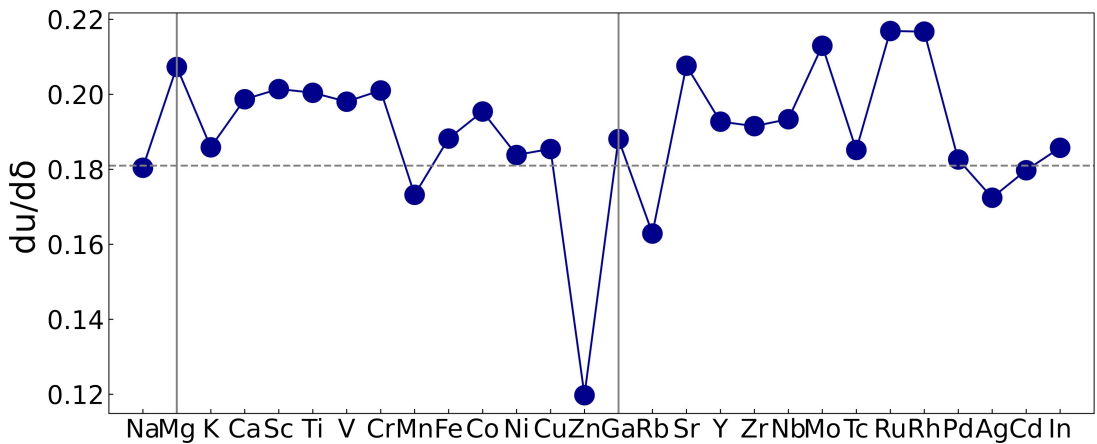

**Figure S5.** $\frac{du}{d\delta}$ of $TM_{0.0625}Al_{0.9375}N$. The grey dashed line denotes the corresponding value in the pure AlN.

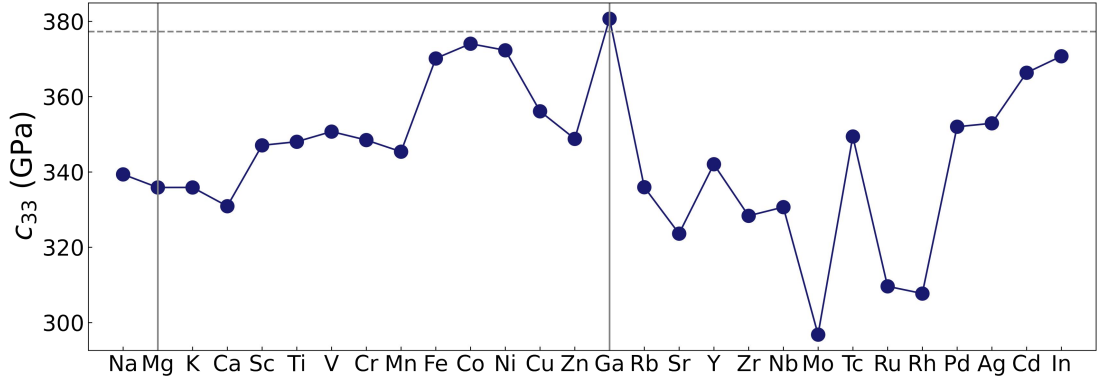

**Figure S6.** $c_{33}$ of $TM_{0.0625}Al_{0.9375}N$. The grey dashed line denotes the corresponding value in pure AlN.

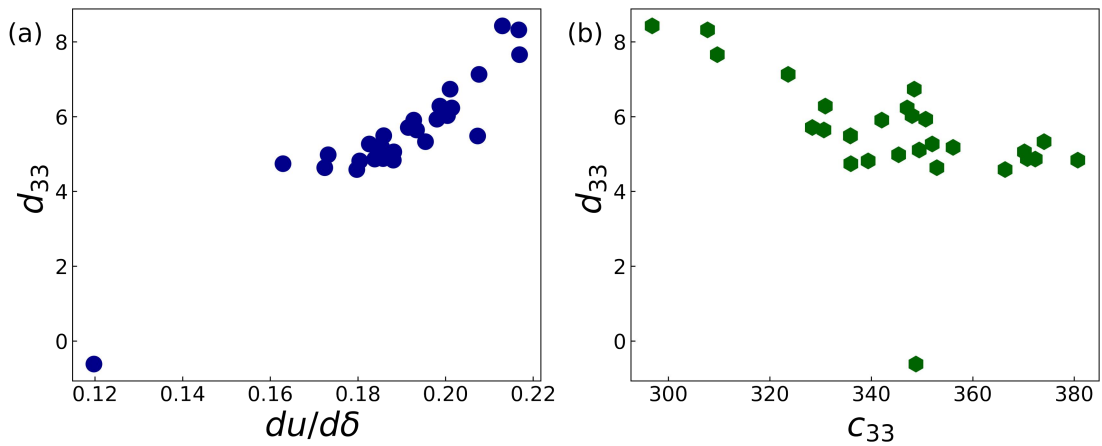

**Figure S7**. a) relationship between $\dfrac{du}{d\delta}$ and $d_{33}$ and b) relationship between $c_{33}$ and $d_{33}$ in $TM_{0.0625}Al_{0.9375}N$.

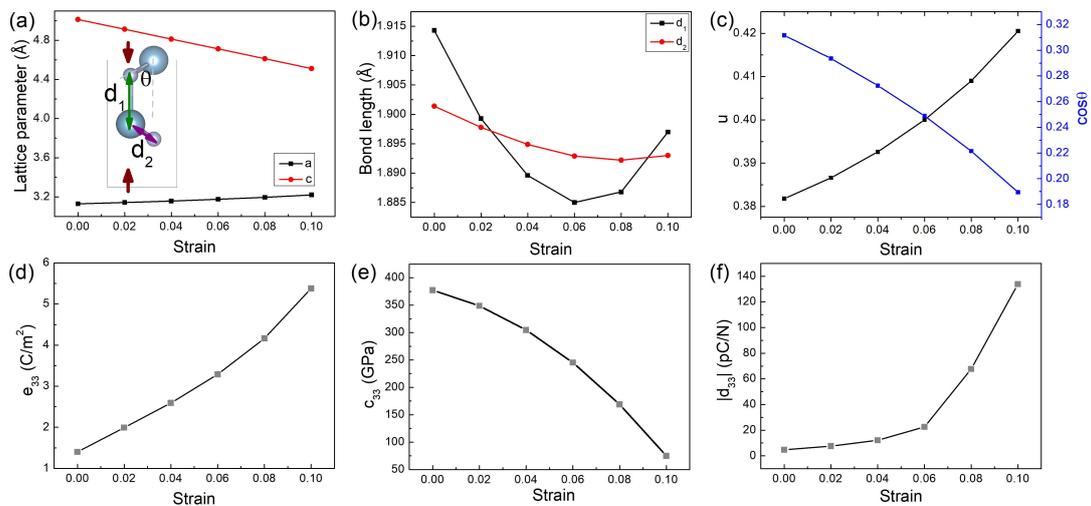

**Figure S8.** a) *a* and *c* of the AlN unit cell under compressive strain along the *z*-axis. The inset shows the AlN unit cell, where the Al-N bond parallel to the *z*-axis is denoted as $d_1$, while the bond with an angle $\theta$ to the $z$-axis is denoted as $d_2$; b) variations in the Al-N bond lengths in the $d_1$ and $d_2$ types under compressive strain; c) variations in $u$ and $cos\,\theta$ under compressive strain; d)-f) variations in $e_{33},\ c_{33}$ and $d_{33}$ respectively under compressive strain.

On the basis of the wurtzite unit cell of AlN shown in Figure S8(a), the internal parameter could be calculated as:

$$u = \frac{d_1}{2(d_1 + d_2\cos\theta)} \tag{S1}$$

From Figure S8, the bond lengths of Al-N show slight variations under compressive strain. $d_1$ (parallel to the $z$-axis) varies more than $d_2$ (with an angle $\theta$ to the $z$-axis), and the magnitude of variation in $d_1$ is within 1.52% under compressive strain up to 0.1. On the other hand, the internal parameter $u$ and the value of $\cos\theta$ vary more significantly, with the magnitudes of variation under a compressive strain of 0.1 at 10.1% and 39.2% respectively. Since the bond lengths of $d_1$ and $d_2$ vary slightly under uniaxial compressive strain, it is reasonable to assume that they are constant so as to simplify the problem. Furthermore, the variations in $u$ and $c$ could be calculated as follows:

$$\begin{aligned}\Delta u &= \frac{d_1}{2(d_1 + d_2\cos\theta')} - \frac{d_1}{2(d_1 + d_2\cos\theta)} \\ &= \frac{d_1 d_2(\cos\theta - \cos\theta')}{2(d_1 + d_2\cos\theta')(d_1 + d_2\cos\theta)}\end{aligned} \tag{S2}$$

$$\begin{aligned}\Delta c &= 2(d_1 + d_2\cos\theta') - 2(d_1 + d_2\cos\theta) \\ &= 2d_2(\cos\theta' - \cos\theta)\end{aligned} \tag{S3}$$

where $\theta$ and $\theta'$ denote the angles under two different uniaxial strains along the $z$-axis. Furthermore, $du/d\delta$ could be estimated as follows:

$$\begin{aligned}\frac{du}{d\delta} &= c_0\frac{du}{dc} \\ &\approx c_0\frac{\Delta u}{\Delta c} \\ &= -\frac{c_0 d_1}{4(d_1 + d_2\cos\theta')(d_1 + d_2\cos\theta)}\end{aligned} \tag{S4}$$

From Equation (S4), $\dfrac{du}{d\delta}$ could be larger when $cos\theta$ is smaller. In other words, when $u$ approaches 0.5 ( $\cos\theta = 0$ ), $\dfrac{du}{d\delta}$ and the corresponding piezoelectric constants $e_{33}$ and $d_{33}$ could increase significantly. This statement could also be verified by considering the piezoelectric constants shown in Figure S8. Additionally, piezoelectric performance under high pressure could also be an interesting topic.

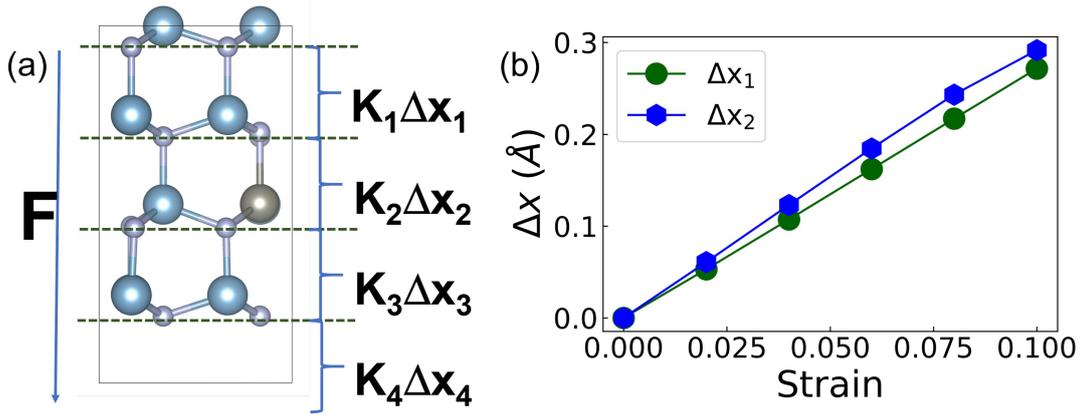

**Figure S9.** a) schematic diagram of a series spring model for the TM-alloy AlN; b) changes in $\Delta x_1$ and $\Delta x_2$ in $Zn_{0.0625}Al_{0.9375}N$ under compressive uniaxial strain.

In order to understand the variation in $\frac{du}{d\delta}$ in Zn- and Rb-alloy AlN, a series spring model is proposed for the TM-alloy configuration which is shown in Figure S9(a). Under uniaxial strain, each sublayer could bear a force $F$. Thus, we have the following equation:

$$K_1 \Delta x_1 = K_2 \Delta x_2 = K_3 \Delta x_3 = K_4 \Delta x_4 = F$$
$$\Delta x_1 + \Delta x_2 + \Delta x_3 + \Delta x_4 = \Delta c \tag{S5}$$

where $K_i \, (i = 1, 2, 3, 4)$ denotes the spring coefficient in each sublayer shown in Figure S9(a), which could be proportional to the bond strength in the sublayer. $\Delta x_i \, (i = 1, 2, 3, 4)$ represents the change in thickness of the sublayer under the uniaxial strain $\Delta c / c$. According to Equation S5, the ratio of $\Delta x_1$ to $\Delta x_2$ satisfies the following equation:

$$\frac{\Delta x_1}{\Delta x_2} = \frac{K_2}{K_1} \tag{S6}$$

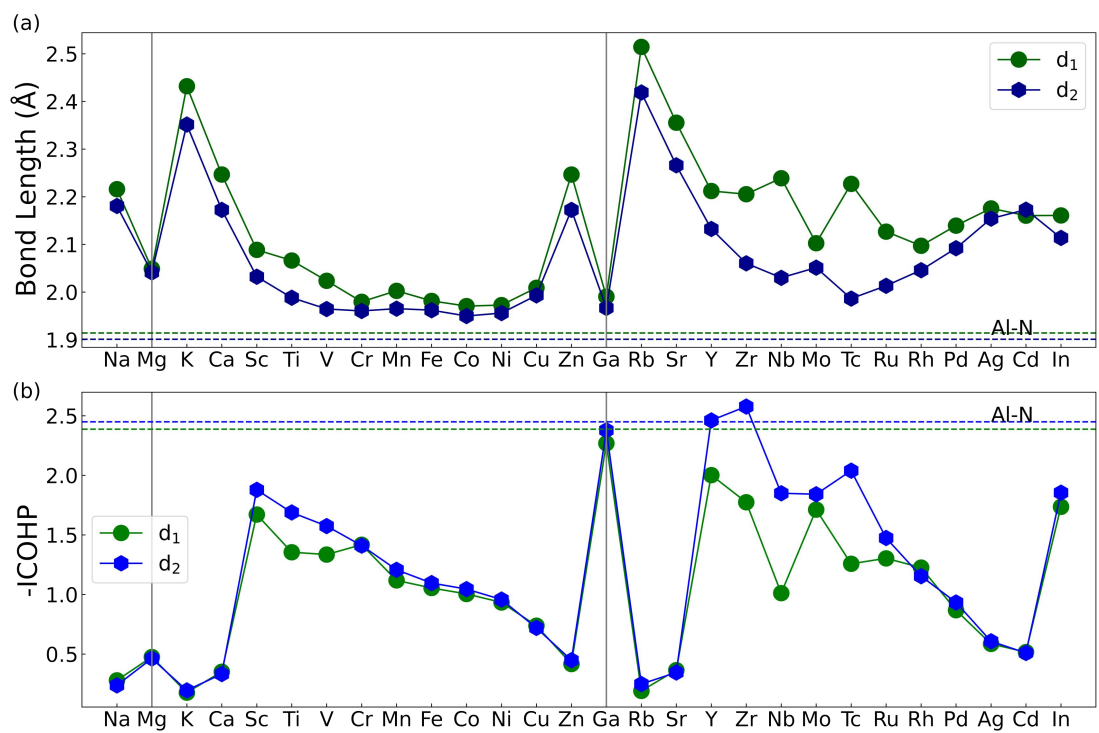

**Figure S10.** a) TM-N bond lengths in $d_1$ (parallel to the $z$-axis) and $d_2$ (with an angle $\theta$ to the $z$-axis) respectively; (b) bond strengths of the TM-N bonds, where the green circles and blue hexagons denote the integrated -COHP values of the TM-N bonds in $d_1$ and $d_2$ respectively, and the green and blue dashed lines denote the corresponding values of Al-N in pure AlN.

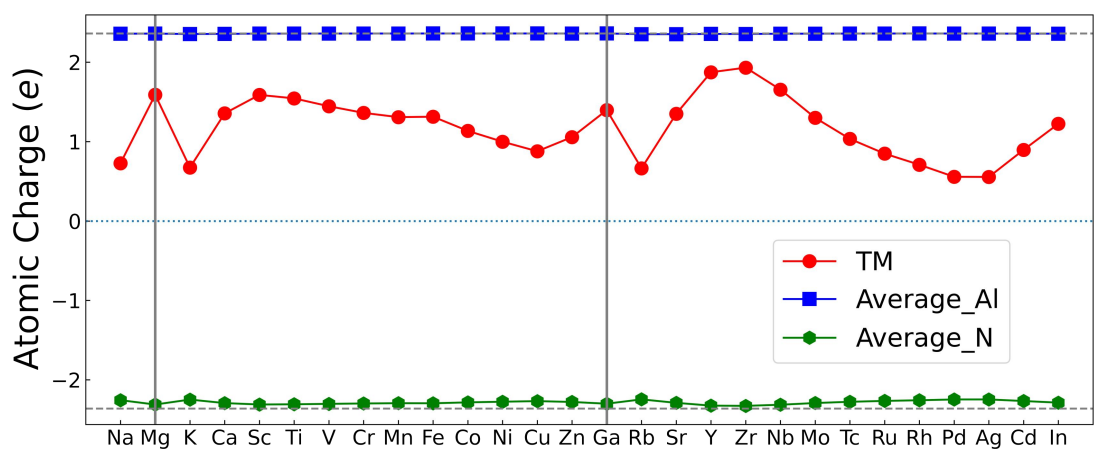

**Figure S11.** Atomic charges in the TM-alloy AlN. The grey dashed lines show the values of Al and N in pure AlN.

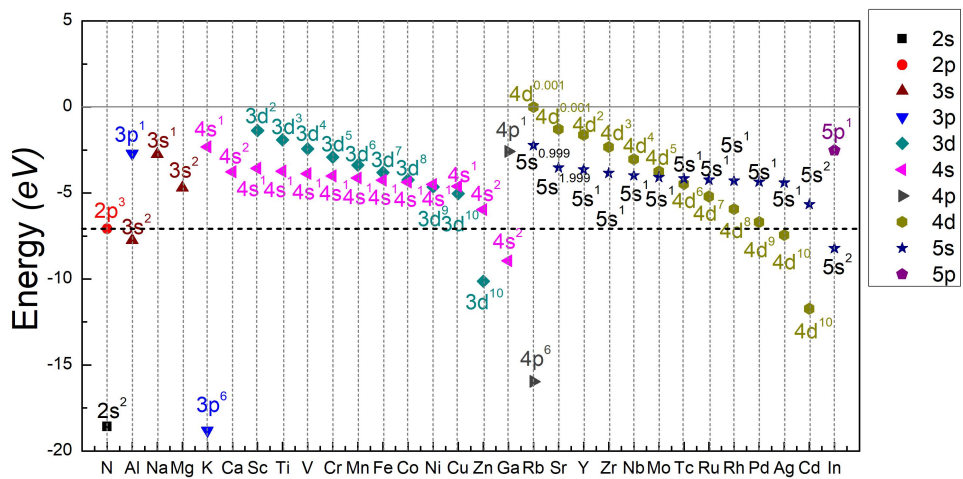

**Figure S12.** Energy, type and occupation of the valence orbitals in N, Al and TM. The data are from the pseudopotential files of VASP.

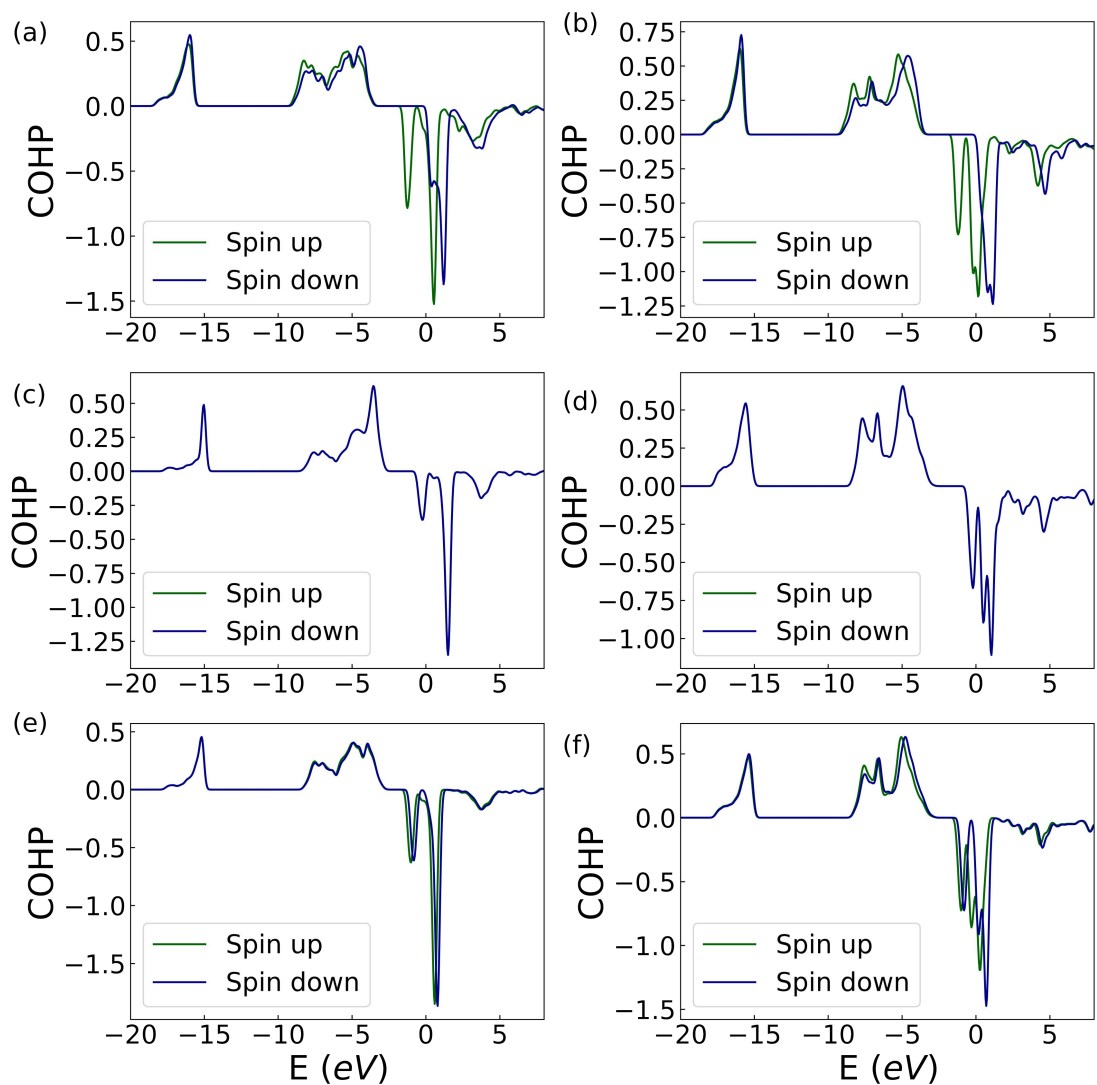

**Figure S13**. COHP of Mo-N in the a) $d_1$ and b) $d_2$ types in Mo$_{0.0625}$Al$_{0.9375}$N; the COHP of Tc-N in the c) $d_1$ and d) $d_2$ types in Tc$_{0.0625}$Al$_{0.9375}$N; and e) and f) are the corresponding COHP of Ru-N in Ru$_{0.0625}$Al$_{0.9375}$N. The COHP from spin up and spin down are denoted in dark green and blue lines respectively.

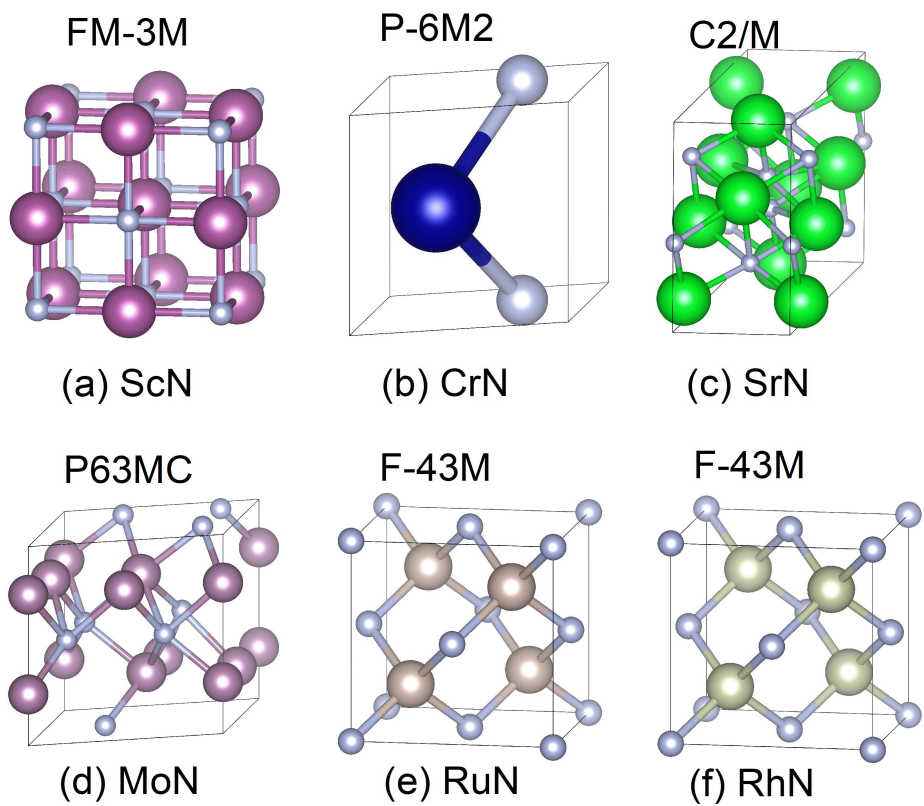

**Figure S14**. Stable structures for a) ScN, b) CrN, c) SrN, d) MoN, e) RuN and f) RhN. Corresponding space groups for these structures are also given.

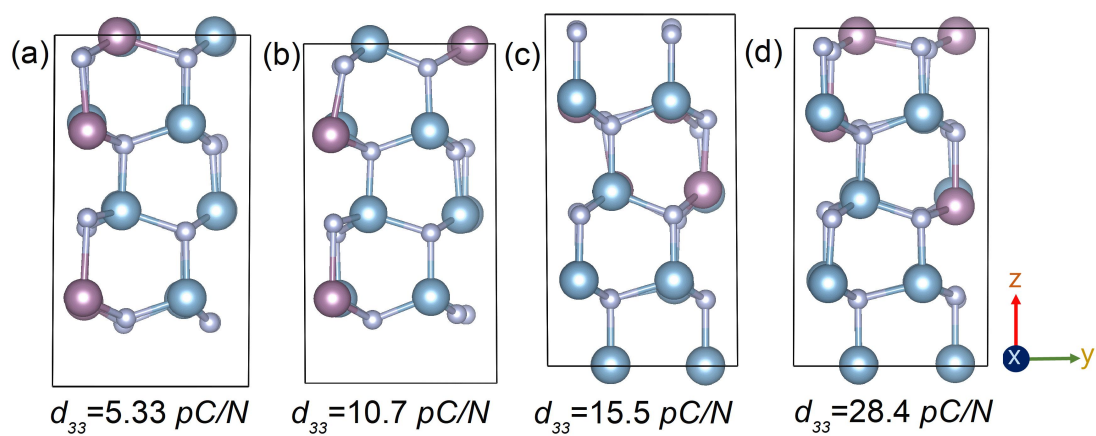

**Figure S15.** Some relaxed SQSs of Mo$_{0.250}$Al$_{0.750}$N with markedly different $d_{33}$ values based on the 2×2×2 AlN supercells and GGA+U; a)-d) SQSs with $d_{33}$ values of 5.33, 10.7, 15.5 and 28.4 *pC/N* respectively.

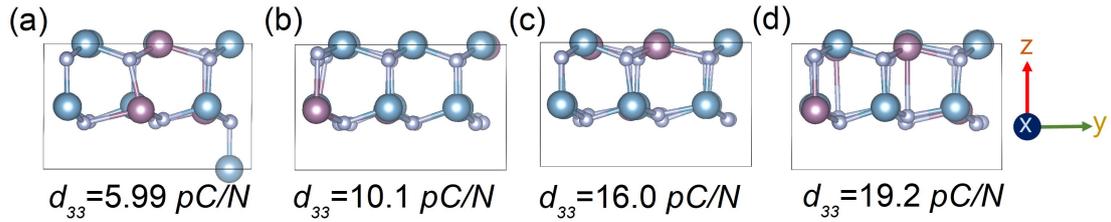

**Figure S16.** Some relaxed SQSs of Mo$_{0.222}$Al$_{0.778}$N with markedly different $d_{33}$ values based on the 3×3×1 AlN supercells and GGA+U; a)-d) SQSs with $d_{33}$ values of 5.99, 10.1, 16.0 and 19.2 *pC/N* respectively.

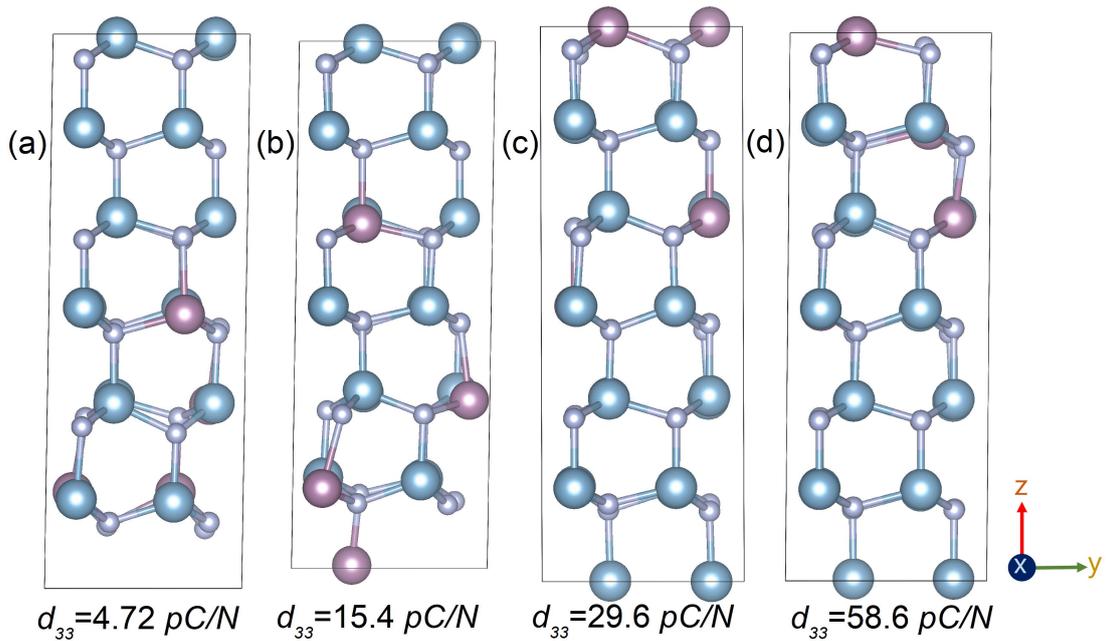

**Figure S17.** Some relaxed SQSs of Mo$_{0.167}$Al$_{0.833}$N with markedly different $d_{33}$ values based on the 2×2×3 AlN supercells and GGA+U; a)-d) SQSs with $d_{33}$ values of 4.72, 15.4, 29.6 and 58.6 *pC/N* respectively.

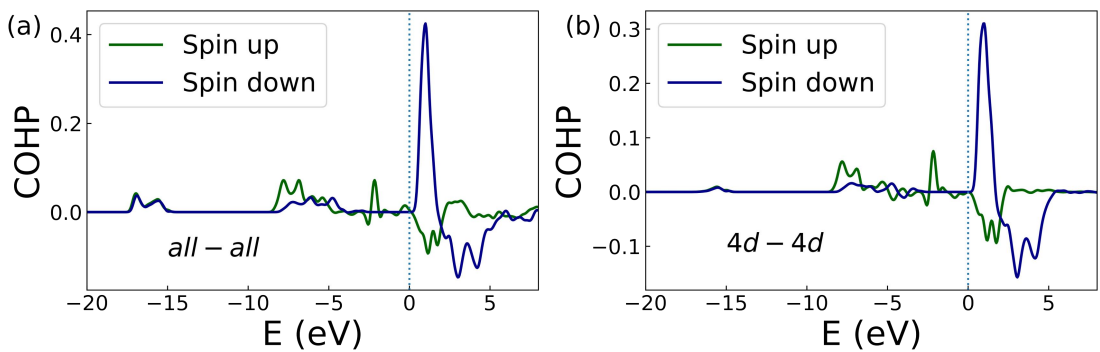

**Figure S18.** a) COHP of the Mo-Mo dimer in Figure S15(a); b) COHP between the 4*d* orbitals of the Mo-Mo dimer. The dark green and dark blue lines show the spin up and spin down states.